\documentclass[aps,prd,reprint,showpacs,floatfix,longbibliography,nofootinbib,superscriptaddress]{revtex4-1}

\usepackage{tikz,graphicx,booktabs,multirow,array,microtype,mathtools,float}

\usepackage{standalone} 
\usepackage{centernot}

\graphicspath{{figures/}}
\usepackage{enumerate}   
\extrafloats{200}
\usepackage[colorlinks=true,backref=false, linktocpage=true,
citecolor=blue,
urlcolor=blue,linkcolor=blue,
pdfpagemode=UseOutlines]{hyperref}
\newcolumntype{C}{>{$}c<{$}}
\usepackage{dcolumn}

\usepackage{anyfontsize}
\usepackage[shortlabels]{enumitem}

\usepackage{amsfonts}
\AtBeginDocument{
\heavyrulewidth=.08em
\lightrulewidth=.05em
\cmidrulewidth=.03em
\belowrulesep=.65ex
\belowbottomsep=0pt
\aboverulesep=.4ex
\abovetopsep=0pt
\cmidrulesep=\doublerulesep
\cmidrulekern=.5em
\defaultaddspace=.5em
}

\newcommand{\langl}{\begin{picture}(4.5,7)
\put(1.1,2.5){\rotatebox{60}{\line(1,0){5.5}}}
\put(1.1,2.5){\rotatebox{300}{\line(1,0){5.5}}}
\end{picture}}
\newcommand{\rangl}{\begin{picture}(4.5,7)
\put(.9,2.5){\rotatebox{120}{\line(1,0){5.5}}}
\put(.9,2.5){\rotatebox{240}{\line(1,0){5.5}}}
\end{picture}}

\usepackage{amsmath} 
\usepackage{dsfont}  
\usepackage{bm}      

\def\iden{\mathds{1}}

\def\beq{\begin{equation}}
\def\eeq{\end{equation}}
\def\beqs#1\eeqs{\beq\begin{split} #1 \end{split}\eeq}

\long\def\comment#1{}

\makeatletter 
    
\renewcommand\onecolumngrid{
\do@columngrid{one}{\@ne}%
\def\set@footnotewidth{\onecolumngrid}
\def\footnoterule{\kern-6pt\hrule width 1.5in\kern6pt}%
}

\renewcommand\twocolumngrid{
        \def\footnoterule{
        \dimen@\skip\footins\divide\dimen@\thr@@
        \kern-\dimen@\hrule width.5in\kern\dimen@}
        \do@columngrid{mlt}{\tw@}
}%

\makeatother    

\begin{document}

\title{
Charged kaon electric polarizability from four-point functions in lattice QCD 
}
\author{Shayan~Nadeem}
\email{shayan\_nadeem1@baylor.edu}
\affiliation{Department of Physics and Astronomy, Baylor University, Waco, Texas 76798, USA}
\author{Walter Wilcox}
\email{walter\_wilcox@baylor.edu}
\affiliation{Department of Physics and Astronomy, Baylor University, Waco, Texas 76798, USA}
\author{Frank~X.~Lee}
\email{fxlee@gwu.edu}
\affiliation{Physics Department, The George Washington University, Washington, DC 20052, USA}

\begin{abstract}

We present a lattice QCD calculation of the electric polarizability of the charged kaon using a four-point function approach, which is the Euclidean analog of low-energy Compton scattering. In the case of the charged kaon, the polarizability is separated into an elastic term, determined from the charge radius extracted via the kaon electromagnetic form factor, and an inelastic term obtained from the time-integrated difference of four-point correlation functions. Our study employs 500 configurations of Wilson quenched $24^3\times 48$ lattices, and we compute connected diagrams as a proof of principle. From this analysis we obtain a charged kaon electric polarizability of
$\alpha_E = (1.682\pm  0.523)\times 10^{-4}~\mathrm{fm}^3$ and a squared
charge radius $r_E^2 = 0.3303 \pm 0.0028~\mathrm{fm}^2$ after extrapolation
to the physical pion mass. The quoted uncertainties include statistical
errors and, for $\alpha_E$, the $\bm q^2 \to 0$ extrapolation; they do not
include systematic effects from the quenched approximation, omitted
disconnected diagrams, finite volume, or the single lattice spacing, which
may be comparable in size. The results at the simulated masses should
therefore be regarded as the primary outcome, with the physical-point
values serving as an indicative extrapolation. The study demonstrates the applicability of the four-point function framework to strange mesons, extends previous four-point function polarizability studies, and provides a foundation for future calculations with increased statistics, dynamical fermions, and improved control of systematic uncertainties.

\end{abstract}
\maketitle

\twocolumngrid

\section{Introduction}
\label{sec:intro} 
The electric polarizability of hadrons characterizes their response to an external electromagnetic field and provides insight into their internal quark and gluon structure. In the low-energy expansion of Compton scattering, the polarizability governs the second-order response of the hadron, complementing static observables such as charge radii and magnetic moments. Precision determinations of polarizabilities are therefore an important benchmark for QCD in the non-perturbative regime and a testing ground for effective theories such as chiral perturbation theory (ChPT).  

The standard method for determining hadronic polarizabilities is the background field approach, which has been extensively employed in calculations of dipole polarizabilities~\cite{Fiebig:1988en,Lujan:2016ffj,Lujan:2014kia,Freeman:2014kka,Freeman:2013eta,Tiburzi:2008ma,Detmold:2009fr,Alexandru:2008sj,Lee:2005dq,Lee:2005ds,Engelhardt:2007ub,Bignell:2020xkf,Deshmukh:2017ciw,Bali:2017ian,Bruckmann:2017pft,Parreno:2016fwu,Luschevskaya:2015cko,Chang:2015qxa,Detmold:2010ts}. Extensions of this framework to higher-order polarizabilities have also been developed~\cite{Davoudi:2015cba,Engelhardt:2011qq,Lee:2011gz,Detmold:2006vu}. In this approach, polarizabilities are extracted from energy shifts obtained via two-point correlation functions, making the overall setup conceptually straightforward. Nevertheless, several distinct challenges arise in practical implementations.

First, the external fields must be sufficiently weak to remain within the regime of linear response. Consequently, the induced energy shifts are extremely small compared to the hadron mass—typically at the level of one part in a million, depending on the field strength. In practice, this difficulty has been addressed by exploiting strong statistical correlations between correlation functions computed with and without the background field.

Second, the implementation of uniform background fields on a finite periodic lattice introduces discontinuities at the boundaries. This issue has largely been mitigated through the use of quantized field strengths consistent with periodic boundary conditions, or alternatively by adopting Dirichlet boundary conditions.

Third, and most importantly for charged states, the presence of an external field induces nontrivial dynamics unrelated to intrinsic structure. A charged hadron accelerates in an electric field and forms Landau levels ~\cite{Bignell_2020,He:2021eha} in a magnetic field. These effects must be disentangled from the deformation energy that defines the polarizability. As a result, the majority of background field calculations have focused on neutral hadrons. For charged hadrons, the two-point correlation function does not exhibit simple single-exponential behavior at large Euclidean times. In Ref.~\cite{Detmold:2009dx}, a relativistic propagator for a charged scalar particle was employed to demonstrate how lattice data for charged pions and kaons can be analyzed in this context. This framework was further refined in Ref.~\cite{niyazi2021charged}, where an effective charged scalar propagator was constructed to exactly match the lattice discretization used in generating the QCD gauge configurations. In addition, a novel fitting strategy was introduced in which the $\chi^2$ function incorporates both the real and imaginary components of the correlator while maintaining invariance under gauge transformations of the background field.

For the pion, polarizabilities have long been the focus of both theoretical and experimental efforts. ChPT provides systematic predictions, and recent lattice QCD studies have begun to achieve direct determinations with controlled uncertainties. The kaon, however, remains less explored. Experimentally, kaon polarizabilities are difficult to access due to the absence of stable kaon targets. This makes the kaon an especially interesting system in which to study the interplay between light and strange quark dynamics.  

In this work, we investigate an alternative strategy based on four-point correlation functions in lattice QCD. Rather than introducing classical background fields, electromagnetic currents are inserted explicitly and coupled to the quark fields, thereby generating interactions to all orders in the electromagnetic coupling. This framework is fully general, placing neutral and charged hadrons on the same theoretical footing, and is particularly advantageous for charged states where background-field methods encounter additional complications. The primary cost of this approach is the increased computational cost and complexity associated with the evaluation of four-point functions. While such correlators have been employed to probe a variety of hadronic structure observables~\cite{Liang:2019frk,Liang_2020a,Fu_2012,Alexandrou_2004,Bali_2018,bali2021double}, comparatively little attention has been devoted to their application in the determination of polarizabilities.
\begin{figure}[t]
\centering
\includegraphics[scale=0.4]{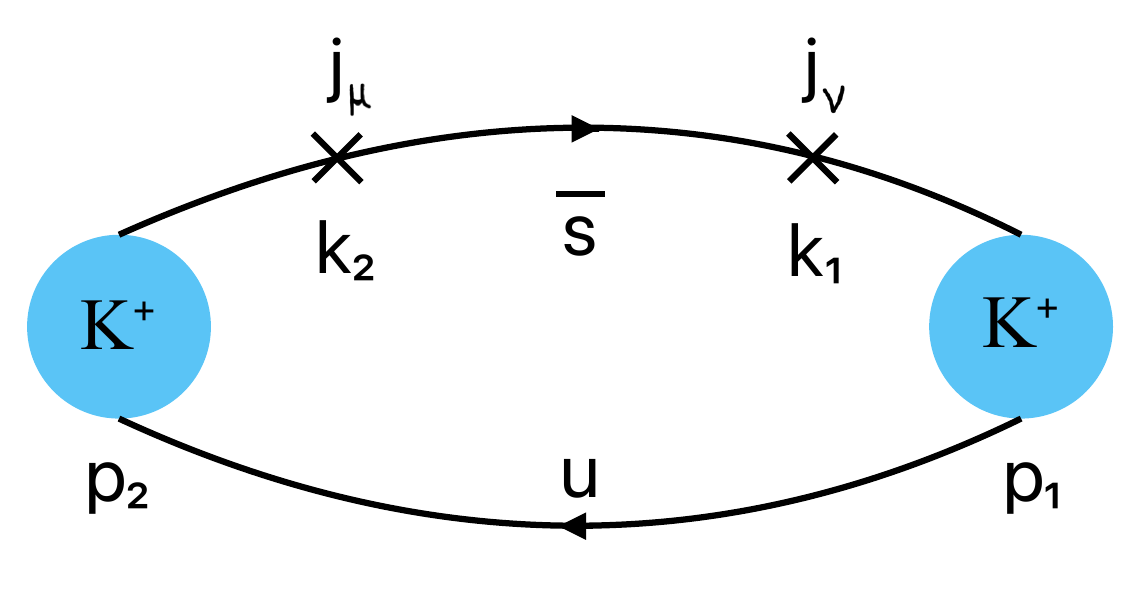}
\caption{Pictorial representation of the four-point function in Eq.\eqref{eq:kaon4pt} for $K^+$. Time flows from right to left and the four-momentum conservation is $p_2 +k_2 = k_1+p_1$. It should be noted that the current densities are applied on the anti-strange quark line. It is also possible to have the current densities on the up quark line or on both. These will be considered as separate diagrams and hence will be evaluated separately. See Fig.~\ref{fig:diagram-4pt1} for the complete list of diagrams after Wick contraction.}
\label{fig:4pdiag}
\end{figure}

Recently, the first lattice determination of the charged pion electric polarizability using four-point functions was reported~\cite{pion4p}. That study demonstrated the feasibility of a method based directly on the Euclidean counterpart of Compton scattering: a meson at rest is probed via two insertions of the electromagnetic current, and the resulting four-point correlation function is analyzed in terms of elastic and inelastic contributions. The charged pion calculation produced results consistent with expectations from chiral effective theory, and established the framework for applying the same methodology to other mesons. Additionally, recent lattice studies have extended this approach to the charged pion magnetic polarizability~\cite{pionmag} and to the polarizabilities of the neutral pion~\cite{neutralpion}. A closely related four-point-function approach has also been applied to the
nucleon electric polarizabilities directly at the physical pion
mass~\cite{Wang2024nucleon}, where the explicit inclusion of $N\pi$
intermediate states was found to be essential for reproducing the
experimental values.

Experimentally, hadronic polarizabilities are predominantly accessed through low-energy Compton scattering. On the theoretical side, a broad range of frameworks has been developed to describe the underlying dynamics. These include quark confinement models~\cite{PhysRevD.45.1580}, the Nambu–Jona-Lasinio (NJL) model~\cite{Dorokhov:1997rv,PhysRevD.40.1615}, the linear sigma model~\cite{BERNARD198816}, dispersion relation analyses~\cite{Lvov:1993fp,PhysRevC.64.015203,PhysRevC.81.029802,Filkov2017Dipole,Stamen:2022uqh,Stamen:2024ocm}, chiral perturbation theory (ChPT)~\cite{Moinester:2019sew,Lensky:2009uv,Hagelstein:2020vog}, and chiral effective field theory (EFT)~\cite{McGovern:2012ew,Griesshammer:2012we}. Comprehensive reviews of hadron polarizabilities are available in Refs.~\cite{moinester2022pionpolarizability2022status,Moinester:2019sew,Griesshammer:2012we}.  

In this paper, we extend the four-point function approach to the kaon. The charged kaon analysis parallels the pion case: the electric polarizability is separated into an elastic term, determined by the kaon charge radius, and an inelastic term extracted from the time-integrated four-point function. Our study presents a proof-of-principle calculation of the charged kaon polarizability. The same analysis strategy and correlator fits used in the pion study are employed here, and we present analogous plots illustrating the separation of elastic and inelastic components. Although carried out on modest ensembles with limited statistics, our results demonstrate the applicability of the four-point function framework to strange mesons and lay the groundwork for future studies with dynamical fermions, increased precision, and systematic control. Additionally, preliminary results for this study were presented at the Lattice 2024 conference~\cite{Nadeem:2025wjf}.

\section{Methodology}

\subsection{Four-Point Function Framework}

We compute the electric polarizability of the charged kaon using lattice QCD four-point correlation functions, analogous to the approach employed for the charged pion~\cite{pion4p}. The expression for the electric polarizability derived in Ref.~\cite{4point} can be straightforwardly adapted to the case of the charged kaon. With the appropriate replacement of hadronic parameters, the resulting expression for the charged kaon electric polarizability consists of an elastic term, determined by the charge radius, and an inelastic term involving the difference between the full and elastic four-point functions,
\begin{equation} \alpha_E= {\alpha \langl r_E^2\rangl \over 3m_{K}}+\lim_{\bm q\to 0}{2\alpha \over \bm q^{\,2}} \int_{0}^\infty d t \bigg[Q_{44}(\bm q,t) -Q^{elas}_{44}(\bm q,t) \bigg]. 
\label{eq:kaon4pt} 
\end{equation}

Here, $\alpha=1/137$ represents the fine structure constant. This formula is applied in discrete Euclidean spacetime, though we retain a continuous Euclidean time axis for ease of notation. Special kinematics, known as the zero-momentum Breit frame, are used in the formula to simulate low-energy Compton scattering. The process is illustrated in Fig.~\ref{fig:4pdiag}.

We emphasize that the separation into `elastic' and `inelastic' terms arises from the Born amplitude organizational structure employed in the four-point-function formulation of Refs.~\cite{pion4p,4point}. The physical electric dipole polarizability is the full quantity $\alpha_E$ defined in Eq.~\eqref{eq:kaon4pt}. Accordingly, the term $\alpha\langle r_E^2\rangle/3m_K$ should be viewed as part of the non-Born (contact term) contribution used in the formulation rather than as a separately observable component of the polarizability. The separation into elastic and inelastic terms is therefore best viewed as a calculational decomposition rather than a decomposition of the physical polarizability itself.

\begin{figure}[b!]
\centering
\includegraphics[scale=1.3]{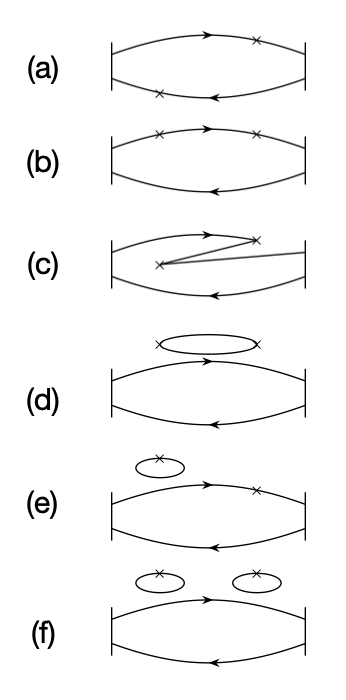}
\caption{Skeleton diagrams of a four-point function contributing
to polarizabilities of a meson: (a) connected insertion: different
flavor, (b) connected insertion: same flavor, (c) connected insertion: same flavor Z-graph, (d) disconnected insertion: single
loop, double current, (e) disconnected insertion: single loop,
(f) disconnected insertion: double loop. In each diagram, flavor
permutations are assumed as well as gluon lines that connect the
quark lines. The zero-momentum kaon interpolating fields are
represented by vertical bars (wall sources). Time flows from right
to left.}
\label{fig:diagram-4pt1}
\end{figure}

The quantity $Q_{44}$ is defined as the temporal--temporal ($\mu=\nu=4$) component of the Fourier-transformed four-point function,
\beqs
Q_{\mu\nu}^{(\text{b,c})}(\bm q,t_2,t_1) \equiv 
\sum_{\bm x_2} 
e^{-i\bm q\cdot \bm x_2} 
D_{\mu\nu}^{(\text{b,c})}(\bm x_2,t_2,t_1).\\
Q_{\mu\nu}^{(\text{a})}(\bm q,t_2,t_1) \equiv 
\sum_{\bm r} 
e^{-i\bm q\cdot \bm r} 
D_{\mu\nu}^{(\text{a})}(\bm r,t_2,t_1).
\label{eq:Qmn}
\eeqs
Here the momentum $\bm q = (0,q_y,q_z)$ and $D_{\mu\nu}^{(\text{b,c})}$ is the four-point correlation function summed over the $x$-component of the vector $\bm x_{1}$. Explicitly,
\beqs
D_{\mu\nu}^{(\text{b,c})}(\bm x_2,t_2,t_1) \equiv
\sum_{\bm x_{1_x}} 
P_{\mu\nu}^{(\text{b,c})}(\bm x_2,\bm x_1,t_3,t_2,t_1,t_0).\\
D_{\mu\nu}^{(\text{a})}(\bm r,t_2,t_1) \equiv
\sum_{\bm x_{1}} 
P_{\mu\nu}^{(\text{a})}(\bm {r+x_1},\bm x_1,t_3,t_2,t_1,t_0).
\label{eq:Qmn2}
\eeqs
The superscripts $(\text{b,c})$ and $(\text{a})$ refer to diagrams $(\text{a})$,$(\text{b})$ , and $(\text{c})$ in Fig.~\ref{fig:diagram-4pt1}. The sum over $\bm x_{1_x}$ in $D_{\mu\nu}^{(\text{b,c})}$ is because we use an extended exactly conserved zero-momentum charge density source in the $x$-direction at time $t_1$. So $\bm x_1 = (x,0,0)$ with $y=0$ and $z=0$ defining the origin in those directions. $P_{\mu\nu}$ here denotes the corresponding four-point correlation function in position space. Explicitly,
\beqs
&P_{\mu\nu}(\bm x_2,\bm x_1,t_3,t_2,t_1,t_0) \equiv \\
&\frac{
\langle \Omega |
\Psi_{{\bar{w}},w}(x_3)\, : j^L_\mu(x_2) j^L_\nu(x_1) : \,
\Psi_{w}^\dagger(x_0)
| \Omega \rangle
}{
\langle \Omega |
\Psi_{{\bar{w}},w}(x_3)\, \Psi_{w}^\dagger(x_0)
| \Omega \rangle
}.
\label{eq:P1}
\eeqs
Here, $\Psi_{\bar{w},w}$ denotes a zero-momentum interpolating field for the kaon, with subscripts specifying the type of wall source (discussed below), and $j^L_\mu$ denotes the lattice discretization of the electromagnetic current, with the product $:j^L_\mu(x_2) j^L_\nu(x_1):$ understood to be normal ordered. The two-point function appearing in the denominator serves to normalize the four-point correlator and remove the overlap factors associated with the interpolating fields. In this study, we use two types of zero-momentum walls: a zero-momentum kaon field defined as,
\begin{equation}
\Psi_{\overline{w}}(t) = \sum_{\bm x} 
\bar{\psi}^{\bar{s},\bar{u}}(\bm x,t)\gamma_5\psi^{u,s}(\bm x,t),\label{zeroW0}
\end{equation}
where $u$ and $s$ denote up and strange quarks respectively, and a zero-momentum quark propagator defined as,
\begin{equation}
\Psi_w(t) = \bar{\psi}^{(0)\bar{s},\bar{u}}(t)\gamma_5\psi^{(0) u,s}(t),\label{zeroW1}
\end{equation}
where
\begin{equation}
\psi^{(0) u,s}(t) \equiv \sum_{\bm x} \psi^{u,s}(\bm x,t). \label{zeroW2}
\end{equation}
We rely on Elitzur’s theorem~\cite{Elitzur}, which guarantees that expectation values of non–gauge-invariant operators vanish, thereby eliminating non–closed quark paths in the four-point function. This procedure is important because it strengthens the signal of the four-point functions we extract.

When the times are well separated (defined by the time limits $t_3\gg t_{1,2} \gg 1$) the correlator is dominated by
the ground state and the structure of the four-point correlator is given as,
\beqs
& P_{\mu\nu}(\bm x_2,\bm x_1,t_3,t_2,t_1,t_0)
\to \\ &\langl K^+(\bm 0) | T(j^{L}_\mu(x_2) j^{L}_\nu(x_1)) |  K^+(\bm 0)\rangl,
 \eeqs
where $T$ denotes time ordering. The current $j^{L}_\mu$ can be applied to any quark line, so generally it is given as,
\begin{equation}
    j^{L}_\mu(x) = q_sj^{L,s}_\mu(x)+q_uj^{L,u}_\mu(x),
\end{equation}
where,
\beqs
 j^{L,u}_\mu (x) &= {\kappa_u} \big[ 
-\bar{\psi}^u(x) (1-\gamma_\mu) U_\mu(x) \psi^u(x+\hat{\mu}) 
\\ &
+
\bar{\psi}^u(x+\hat{\mu}) (1+\gamma_\mu ) U_\mu^\dagger(x) \psi^u(x) 
\big],
\label{eq:current}
\eeqs
and similarly for the $s$ quark.

We employ a different approach to the one used in Ref.~\cite{pion4p}: the calculation here enables the reconstruction of all momenta in the $y$ and $z$ directions from a single set of input propagators, whereas the earlier method treats each momentum independently. The restriction to momentum components in the $y$ and $z$ directions arises from the use of an extended, exactly conserved zero-momentum charge-density source in the $x$-direction at time $t_1$. As a result, the four-point function yields a spatial charge-density distribution in the transverse $(y,z)$ plane, which is subsequently analyzed to extract the desired momentum dependence. This strategy follows the Fourier Reinforcement (FR) technique introduced in Ref.~\cite{Wilcox:1993uq}, which exploits extended sources to improve statistical precision, while limiting the momentum reconstruction to directions transverse to the source extension.
Note that in our setup described in Sect.\ref{sec:2B}, momentum reconstruction is performed entirely in the final analysis, and no explicit momentum projection is imposed during propagator construction. All injected momenta can be reconstructed in this formalism, subject to the FR source transversality condition. We will see that the combination of the FR method and our reconstruction formalism yields a strong numerical signal with a minimal suppression of allowed momentum analysis values.

The four-point function method is based on the Euclidean-space analog of low-energy Compton scattering. Following the standard decomposition, Eq.\eqref{eq:kaon4pt} is expressed as a sum of elastic and inelastic contributions:
\begin{equation}
\alpha_E = \alpha_E^{\mathrm{elastic}} + \alpha_E^{\mathrm{inelastic}}.
\end{equation}
For the charged kaon, the elastic contribution is determined by the kaon electromagnetic form factor $F_K(q^2)$:
\begin{equation}
\alpha_E^{\mathrm{elastic}} \equiv {\alpha \langl r_E^2\rangl \over 3m_{K}},
\end{equation}
where 
\begin{equation}
\langl r_E^2 \rangl \equiv -6 \left. \frac{dF_K(q^2)}{d q^2} \right|_{q^2 \to 0},   
\label{chargeradius}
\end{equation}
is the kaon charge radius squared and $m_K$ is the kaon mass.

On the lattice, $F_K(q^2)$ is extracted by fitting the large-time behavior of the four-point function:
\begin{equation}
Q_{44}^{\mathrm{elas}}(\bm{q},t) = \frac{(E_K + m_K)^2}{4 E_K m_K} |F_K(q^2)|^2 e^{-(E_K - m_K) t},
\label{eq:elas4pt}
\end{equation}
where $E_K = \sqrt{m_K^2 + \bm q^2}$ is the kaon energy and the form factor appears as squared as a function of space-like four-momentum squared $q^2$.

The inelastic contribution is obtained by subtracting the elastic part from the full four-point correlator and integrating over Euclidean time:
\begin{equation}
\alpha_E^{\mathrm{inelastic}} = \lim_{\bm q\to 0} \frac{2 \alpha}{\bm q^2} \int_0^{\infty} dt \, [ Q_{44}(\bm q,t) - Q_{44}^{\mathrm{elas}}(\bm q,t) ].
\end{equation}

For a charged pseudoscalar the charge radius enters the polarizability formula, Eq.~\eqref{eq:kaon4pt}, explicitly through the elastic term $\alpha\langl r_E^2\rangl/3m_K$, whose coefficient is fixed by current
conservation. For a neutral pseudoscalar this elastic contribution vanishes and the term is absent, as was shown for the neutral pion~\cite{neutralpion}. The four-point-function method thereby accounts for the distinction between charged and neutral states automatically. This is a notable advantage over the background-field approach, in which the
acceleration energy of a charged hadron in the applied field is entangled with the polarization energy that defines the polarizability.

Part of computing the four-point function is to evaluate the topologically distinct quark-line diagrams. These diagrams are shown in Fig.~\ref{fig:diagram-4pt1}. The raw correlation functions are discussed in the following sections.
\subsection{Four-point Functions} \label{sec:2B}
We will now individually go over the first three diagrams shown in Fig.~\ref{fig:diagram-4pt1}. In this work, we only consider the connected diagrams. For clarity, in this section we will discuss the flavor-symmetric case where both $u$ and $s$ quark propagators for the kaon have the same mass. For the case where they have different masses, there are actually two combinations for each diagram. We use the term sequential source technique (\lq\lq SST")~\cite{Gattringer:2010zz, Mtter1986LatticeGT, KILCUP1985347} to describe the construction of secondary quark propagators via inversion on a previously computed propagator. 
\subsubsection{Diagram (a)}
\begin{figure}[h]
    \centering  \includegraphics[scale=1.3]{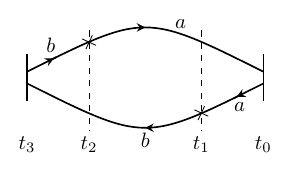}
    \caption{Diagram (a) shown in terms of quark propagators. The quark line labeled \lq\lq $a$" comes from a quark wall source at $t_0$. The quark line labeled \lq\lq $b$" comes from a quark wall source at $t_3$.}
    \label{fig:Diagram a}
\end{figure}

Diagram (a) has a current insertion on each quark line, as shown in Fig.~\ref{fig:Diagram a}.
\begin{figure}[h]
\centering
\includegraphics[scale=0.3]{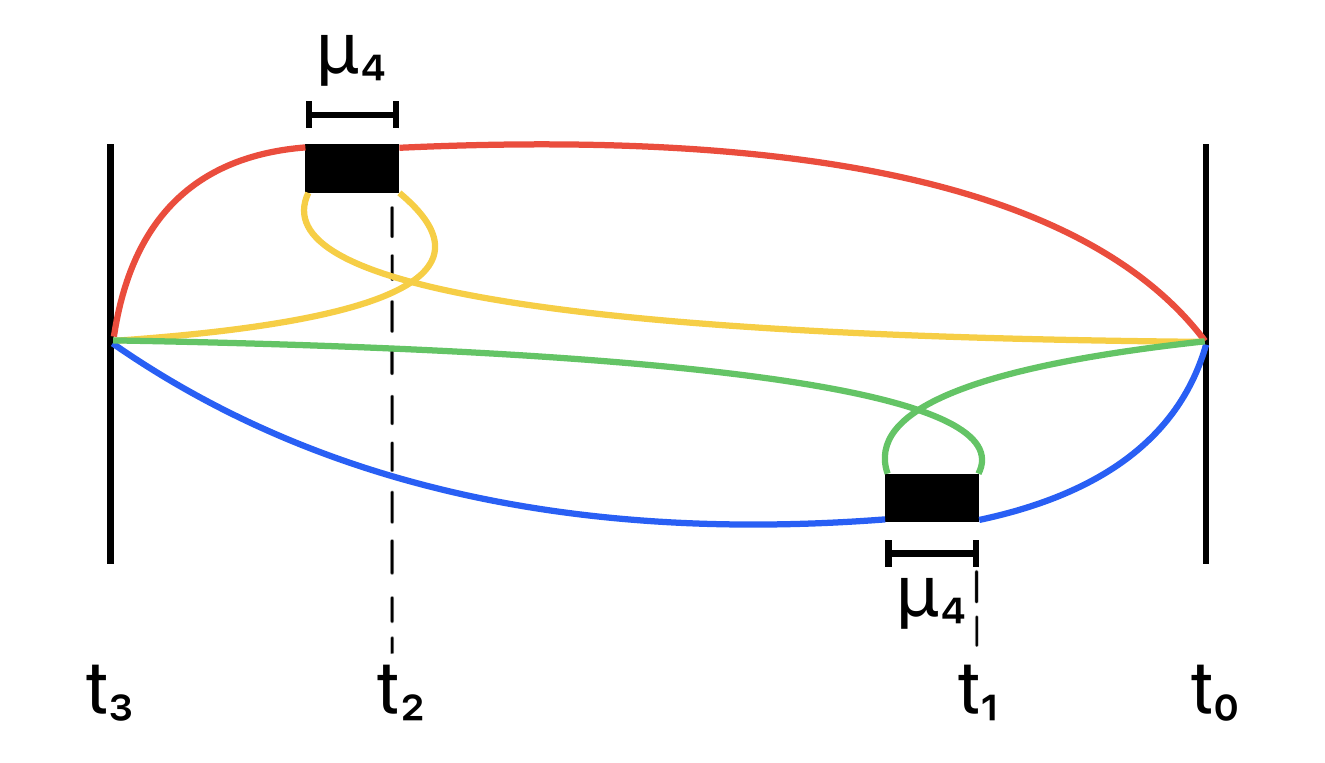}
\caption{Diagram (a) quark flow diagram. The direction of the quark flow is as in Fig.~\ref{fig:Diagram a}.}
\label{fig:diagaflow}
\end{figure}
By following the trace of the loop in diagram (a) in Fig.~\ref{fig:Diagram a} in a clockwise manner, the interaction amplitude is (Ref.~\cite{pion4p}):
\begin{multline}
    D_{44}^{(a)}(\bm r,t_2,t_1) = 
    \sum_{\bm x_1 (\bm x_2=\bm x_1+ \bm r)}\\
 Tr [ ( \gamma_5 S_w(t_0,t_1)(1-\gamma_4){U}_4(t_1,t_1+\mu_4)S_w(t_1+\mu_4,t_3) \\
 +\gamma_5 S_w(t_0,t_1+\mu_4)(1+\gamma_4){U}^\dagger_4(t_1+\mu_4,t_1)S_w(t_1,t_3) )  \\
 \times ( \gamma_5 S_w(t_3,t_2)(1-\gamma_4)U_4(t_2,t_2+\mu_4)S_w(t_2+\mu_4,t_0)  \\
 +\gamma_5 S_w(t_3,t_2+\mu_4)(1+\gamma_4)U^\dagger_4(t_2+\mu_4,t_2)S_w(t_2,t_0)) ].
    \label{eq:amp_a}
\end{multline}
Here $S_w$ is the quark propagator coming from a wall, Eqs.~\eqref{zeroW1} and \eqref{zeroW2}, and the $\bm x_1, \bm x_2$ dependence is implicit in the trace. The standard quark propagator $S(t,t')$ is defined (with implicit spatial, color and spin dependence) as,
\begin{equation}
    S(t,t') \equiv M_q^{-1}(t,\bm {x};t',\bm {x'}),
\end{equation}
where $M_q$ is the Wilson-Dirac quark matrix,
\begin{multline}
M_q (t,t')=
\iden - \kappa_q \sum_{\mu} \Big[(1-\gamma_\mu) U_\mu (t,t+\hat{\mu})+ \\
(1+\gamma_\mu) U_\mu ^\dagger  (t,t-\hat{\mu})\Big],
\label{eq:matW}
\end{multline}
(final time $t$ fixed and position arguments suppressed) where $\kappa_q =1/(2m_q+8)$ is the hopping parameter and $m_q$ the bare quark mass. The space, color and spin dependence of $S(t,t')$ is implicit. So then, using the source Eq.~(\ref{zeroW2}) we define $S_w(t,t')$ as,
\begin{equation}
    S_w(t,t') \equiv \sum_{\bm {x'}} M_q^{-1}(t,\bm {x};t',\bm {x'}).
\end{equation}

The four terms in the amplitude, Eq.~(\ref{eq:amp_a}), come from the four possible paths for the quark lines. This is shown in Fig.~\ref{fig:diagaflow}. The blue and green lines can each be followed by the red and yellow lines, hence giving us the four terms.

The time reversal interchange
\begin{equation}
\begin{matrix}
        t_0 \leftrightarrow t_3, \\
        t_1 \leftrightarrow t_2,  
\end{matrix}
\end{equation}
reverses the direction of the time progression in the diagrams. In other words, time reversal symmetry will interchange the quark and anti-quark in the diagrams. Equivalently, one may change the direction of the meson loop trace from clockwise to counterclockwise. We also have the $\gamma_5$-hermiticity identity (the \lq\lq $\dagger$" acts only in Dirac, color space),
\begin{equation}
    S^{\dagger}(t,t') = \gamma_5 S (t',t) \gamma_5,
\end{equation}
for Wilson-Dirac quark propagators.

\subsubsection{Diagram (b)}

\begin{figure}[h]
\centering
\includegraphics[scale=1.3]{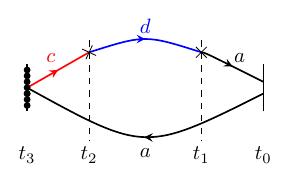}
\caption{Diagram (b) shown in terms of quark propagators. The quark line labeled \lq\lq $a$" comes from a quark wall source at $t_0$. This line propagates to $t_3$ to produce a summed point SST propagator labeled as \lq\lq $c$" and shown as a line with black dots. The \lq\lq $a$" propagator is also used to produce a conserved charge density SST source at $t_1$ labeled as \lq\lq $d$". The \lq\lq $c$" and \lq\lq $d$" quark lines are then sewn together to produce the charge density operator at time $t_2$.}\label{diagramb}
\end{figure}
\begin{figure}[h]
\centering
\includegraphics[scale=0.3]{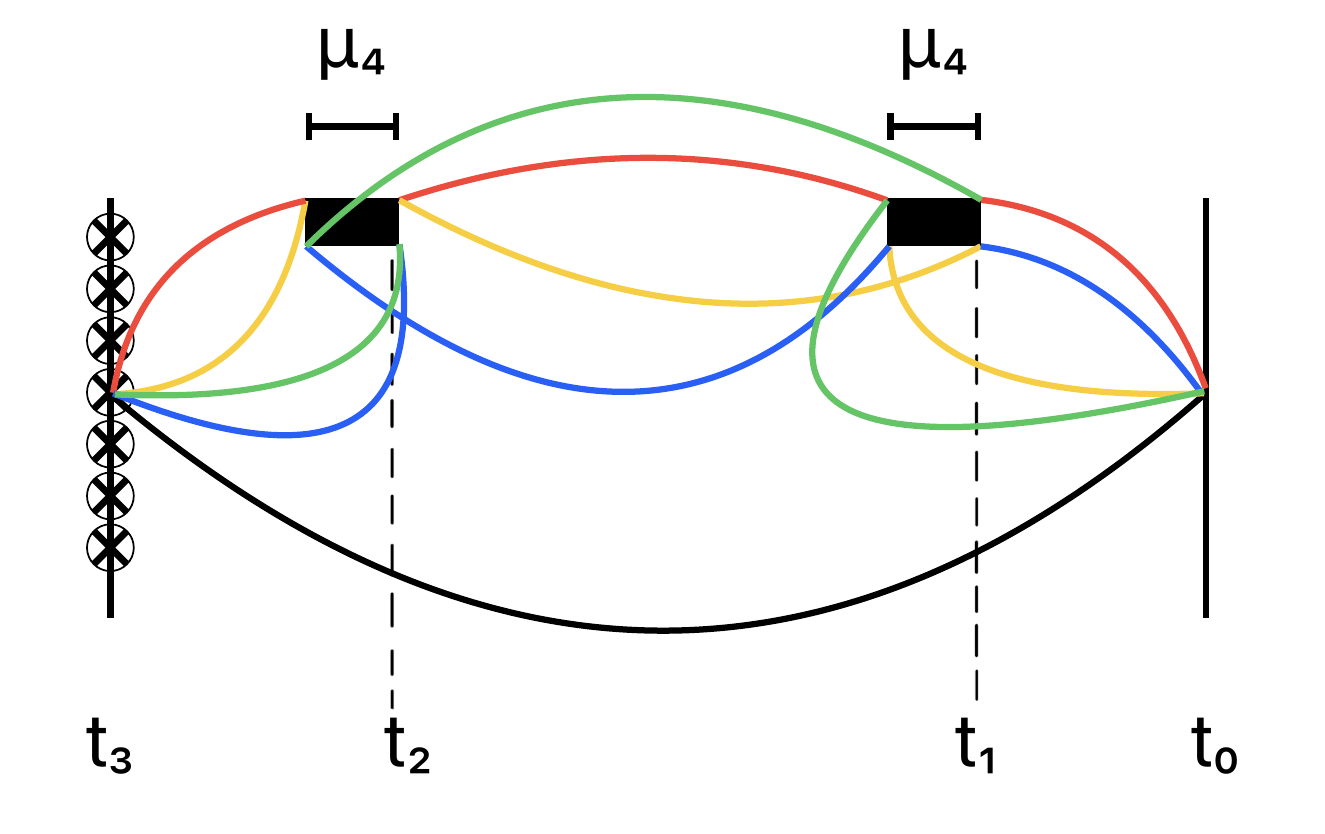}
\caption{Diagram (b) quark flow diagram. The direction of the quark flow is as in Fig.~\ref{diagramb}.}
\label{fig:diagbflow}
\end{figure}

By following the trace of the loop in diagram (b) in Fig.~\ref{diagramb} in the clockwise direction, the interaction amplitude is (Ref.~\cite{pion4p}):
\begin{multline}
D_{44}^{(b)}(\bm x_2,t_2,t_1) =\\\sum_{\bm x_{1_x}, \bm x_3   }
    Tr[\gamma_5 S_{{w}}(t_3,t_0)\gamma_5 S_w(t_0,t_1+\mu_4)(1+\gamma_4)U^\dagger_4 (t_1+\mu_4,t_1)\\
  \times \begin{pmatrix}
        S(t_1,t_2+\mu_4)(1+\gamma_4){U}^\dagger_4(t_2+\mu_4,t_2)S(t_2,t_3) \\
        -S(t_1,t_2)(1-\gamma_4){U}_4(t_2,t_2+\mu_4)S(t_2+\mu_4,t_3)
    \end{pmatrix} \\   
    -\gamma_5 S_{{w}}(t_3,t_0)\gamma_5 S_w(t_0,t_1)(1-\gamma_4)U_4(t_1,t_1+\mu_4)\\
    \times \begin{pmatrix}
        S(t_1+\mu_4,t_2+\mu_4)(1+\gamma_4){U}^\dagger_4(t_2+\mu_4,t_2)S(t_2,t_3) \\
        -S(t_1+\mu_4,t_2)(1-\gamma_4){U}_4(t_2,t_2+\mu_4)S(t_2+\mu_4,t_3)
    \end{pmatrix} ].\label{eq:diagramb}
\end{multline}
The sum over $\bm x_3$ arises from the use of a zero-momentum source applied at time $t_3$, as defined in Eq.\eqref{zeroW0}. The four terms in the amplitude, Eq.\eqref{eq:diagramb}, come from the four possible paths for the quark lines. This is shown in Fig.~\ref{fig:diagbflow} by the different colored lines. 

We next define the SST associated with the kaon source at time $t_3$, which gives rise to the ``$c$'' propagator shown in Fig.~\ref{diagramb},
\begin{equation}
S^{SST,c}(t_f,t_0)\equiv  \sum_{\bm x_3 } S(t_f,t_3)\gamma_5 S_{{w}}(t_3,t_0).
\end{equation}
In addition, we define the SST corresponding to the conserved charge density insertion, which generates the ``$d$'' propagator in Fig.~\ref{diagramb}. At this stage, the sum over the $\bm x_{1_x}$ coordinate of the charge density source is performed.
\begin{multline}
S^{SST,d}(t_0,t_f)\equiv \\ \sum_{\bm x_{1_x}} \{
\gamma_5 S(t_f,t_1+\mu_4)(1+\gamma_4){U}^{\dagger}_4(t_1+\mu_4,t_1)S_w(t_1,t_0) \\
-\gamma_5 S(t_f,t_1)(1-\gamma_4){U}_4(t_1,t_1+\mu_4)S_w(t_1+\mu_4,t_0) \}.
\end{multline}
From this definition it follows that, 
\begin{multline}
\gamma_5 {S^{SST,d}}^{\dagger}(t_0,t_f)= \\ \sum_{\bm x_{1_x}}
\{ S_w(t_0,t_1)(1-\gamma_4){U}_4(t_1,t_1+\mu_4)S(t_1+\mu_4,t_f) \\
-S_w(t_0,t_1+\mu_4)(1+\gamma_4){U}^{\dagger}_4(t_1+\mu_4,t_1)S(t_1,t_f) \}.
\end{multline}
With these definitions, diagram (b) can be expressed in the compact form,
\begin{multline}
D_{44}^{(b)}(\bm x_2,t_2,t_1) =\\
Tr[{S^{SST,d}}^{\dagger}(t_0,t_2)(1-\gamma_4)U_4(t_2,t_2+\mu_4)S^{SST,c}(t_2+\mu_4,t_0)\\
-{S^{SST,d}}^{\dagger}(t_0,t_2+\mu_4)(1+\gamma_4)U_4^{\dagger}(t_2+\mu_4,t_2)S^{SST,c}(t_2,t_0)].
\end{multline}

\subsubsection{Diagram (c)}

\begin{figure}[h!]
\centering
\includegraphics[scale=1.3]{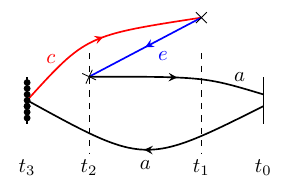}
\caption{Diagram (c) shown in terms of quark propagators. As in diagram (b), the quark line \lq\lq $a$" comes from a quark wall source at $t_0$. This line propagates to $t_3$ to produce a summed point SST source labeled as \lq\lq $c$" and shown as a line with black dots. The quark line \lq\lq $c$" is then used as the SST source at $t_1$ to produce a conserved charge density SST source at $t_1$ labeled as \lq\lq $e$". The \lq\lq $e$" and \lq\lq $a$" quark lines are then sewn together to produce the charge density operator at variable time $t_2$.}\label{diagramc}
\end{figure}
\begin{figure}[h]
\centering
\includegraphics[scale=0.3]{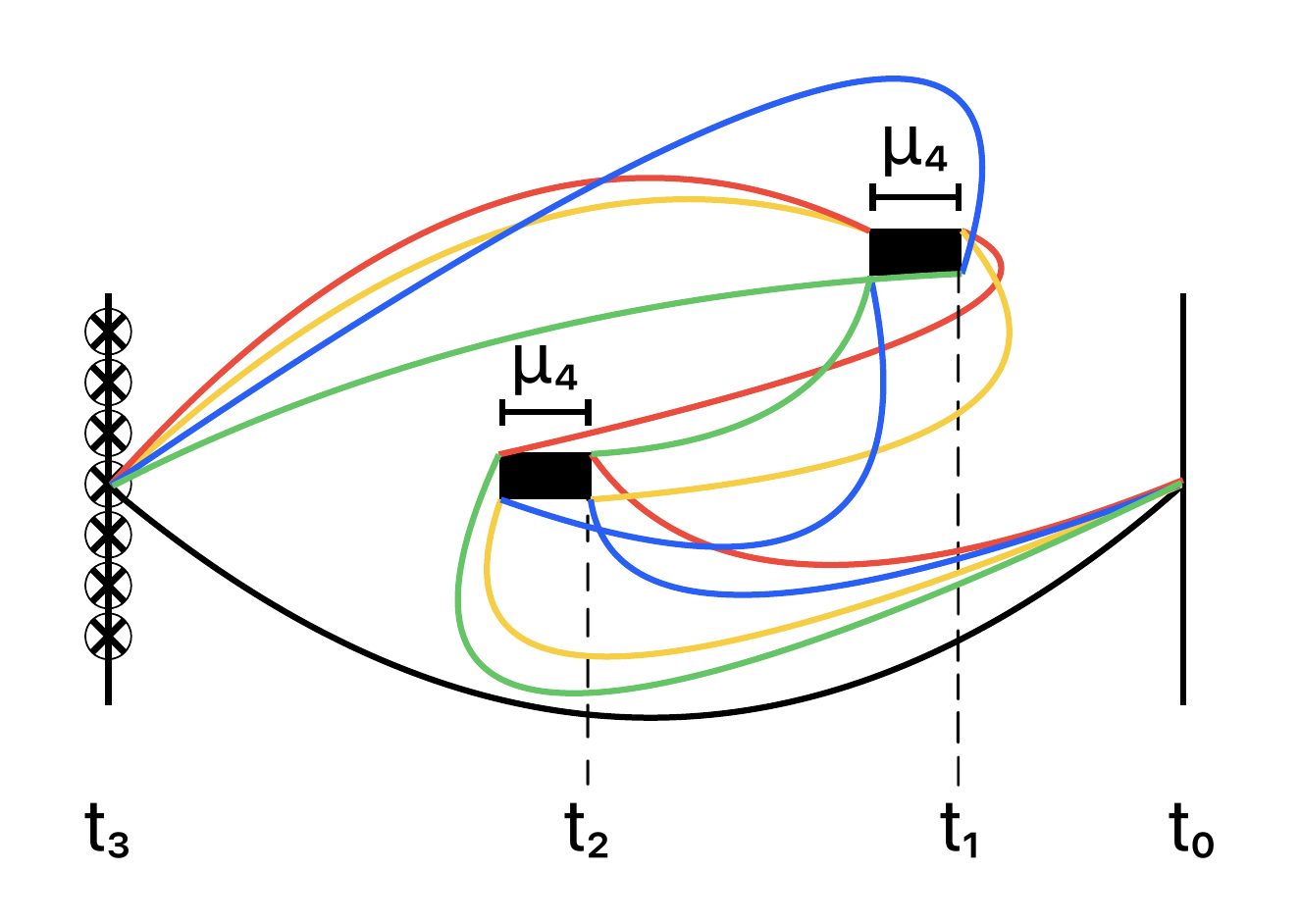}
\caption{Diagram (c) quark flow diagram. The direction of the quark flow is as in Fig.~\ref{diagramc}.}
\label{fig:diagcflow}
\end{figure}
The diagram (c) expression, shown in terms of quark propagators in Fig.~\ref{diagramc}, follows from diagram (b) just by the interchange $t_1 \leftrightarrow t_2$ in Eq.~(\ref{eq:diagramb}). However, we compute it quite differently because we wish to evaluate it on the same lattice time slices as the other diagrams in order to avoid any systematic effects. We use the \lq\lq $c$" SST propagator as the source for the \lq\lq $e$" propagator,
\begin{multline}
S^{SST,e}(t_f,t_0)\equiv \\
\sum_{\bm x_{1_x}}
\{ S(t_f,t_1+\mu_4)(1+\gamma_4){U}^{\dagger}_4(t_1+\mu_4,t_1) S^{SST,c}(t_1,t_0) \gamma_5\\
-S(t_f,t_1)(1-\gamma_4){U}_4(t_1,t_1+\mu_4) S^{SST,c}(t_1+\mu_4,t_0)\gamma_5 \}.
\end{multline}
Then diagram (c) may be written simply as,
\begin{multline}
D_{44}^{(c)}(\bm x_2,t_2,t_1) = \\
Tr[ S_w(t_0,t_2+\mu_4)(1+\gamma_4)U^{\dagger}_4(t_2+\mu_4,t_2){S^{SST,e}}(t_2,t_0) \\
- S_w(t_0,t_2)(1-\gamma_4)U_4(t_2,t_2+\mu_4){S^{SST,e}}(t_2+\mu_4,t_0)].
\label{eq:diagramc}
\end{multline}
The four terms in the amplitude, Eq.\eqref{eq:diagramc}, come from the four possible paths for the quark lines. This is shown in Fig.~\ref{fig:diagcflow} by the different colored lines.

\subsection{Extraction Procedure}
We begin by computing the Euclidean four-point correlation function, $Q_{44}(\bm q,t)$, for a range of Euclidean time separations \(t\) and several values of the spatial momentum transfer \(\bm q\). The kaon is taken to be at rest, while equal and opposite spatial momenta are injected at the two electromagnetic current insertions. The correlation function is evaluated on each gauge ensemble and averaged over equivalent momentum directions to improve statistical precision. This four-point function represents the Euclidean-space analogue of low-energy Compton scattering and constitutes the primary quantity from which the electric polarizability is extracted.

For the charged kaon, the four-point function contains a dominant elastic contribution arising from an intermediate on-shell kaon state. This elastic component, denoted \(Q_{44}^{\mathrm{elas}}(\bm q,t)\), is isolated and fitted using its expected functional form, which depends on the kaon electromagnetic form factor \(F_K(q^2)\). Correlated fits are performed at fixed momentum transfer to extract \(F_K(q^2)\) over the range of accessible four-momentum \(q^2\). The kaon charge radius is then determined from the slope of the form factor at vanishing momentum transfer as shown in Eq.~(\ref{chargeradius}). This quantity fully specifies the elastic contribution to the electric polarizability.

Once the elastic contribution has been determined, it is subtracted from the total four-point correlator to isolate the inelastic component,
$Q_{44}^{\mathrm{inel}}(\bm q,t)= Q_{44}(\bm q,t) - Q_{44}^{\mathrm{elas}}(\bm q,t)$.
The resulting inelastic correlator contains contributions from excited and multiparticle intermediate states and encodes the structure-dependent response of the kaon to an external electric field.

The inelastic correlator is then integrated over the Euclidean time separation \(t\), with careful control of fit ranges to suppress excited-state contamination and finite-time effects. This time-integrated quantity is evaluated for each nonzero value of \(\bm q^2\) and subsequently extrapolated to the \(\bm q^2 \to 0\) limit. This yields the inelastic contribution to the kaon electric polarizability, \(\alpha_E^{\mathrm{inelastic}}\), in accordance with the low-energy expansion of the Compton amplitude.

For the charged kaon, the total electric polarizability is obtained by combining the elastic contribution, determined from the charge radius, with the inelastic contribution extracted from the time-integrated correlator. The final result therefore incorporates both the elastic response associated with the kaon’s charge distribution and the structure-dependent inelastic effects, providing a complete determination of the charged kaon electric polarizability within this framework.
\section{Simulation details and results}
\label{sec:results}
\begin{figure}[h]
\centering
\includegraphics[scale=0.6]{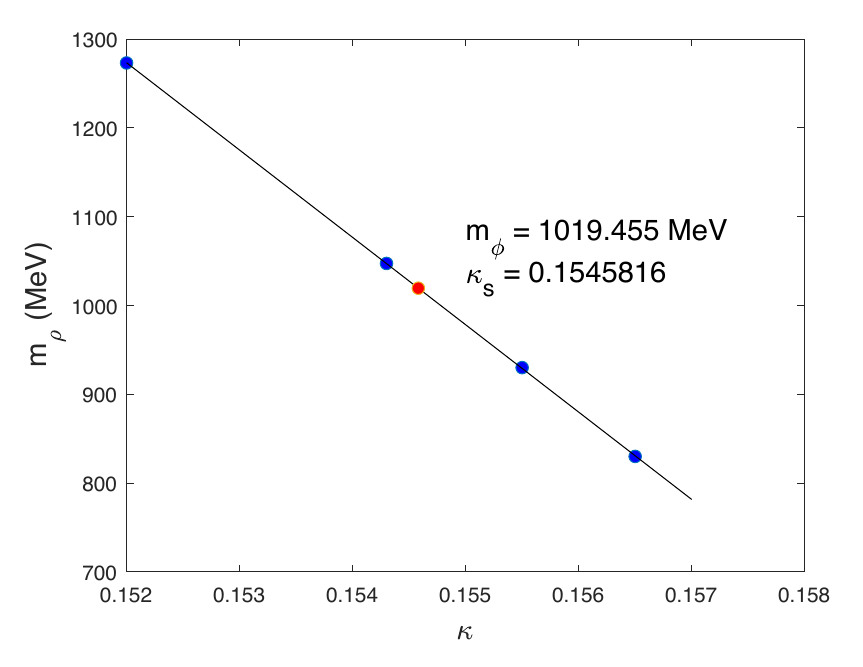} \\
\caption{Data points from Ref.~\cite{pion4p} are shown in blue. The red point indicates the interpolated value of $\kappa$ corresponding to the strange quark, obtained using the $\phi$ meson mass.}
\label{fig:skappa}
\end{figure}

The numerical evaluation of the kaon polarizability is performed using large-scale Monte Carlo simulations designed to sample the path integral of lattice QCD with high statistical fidelity. For the present study, we employ the quenched Wilson gauge action at a coupling of $\beta = 6.0$, together with four kappa values, $\kappa = 0.1543,\; 0.154581,\; 0.1555,\; 0.1565,$
on a lattice of volume $24^3 \times 48$. The corresponding approximate pion masses for these kappas are,
$m_\pi = 800,\; 750,\; 600,\; 370$ MeV. Hereafter, ensembles are denoted by their pion masses. These parameters correspond to a range of light quark masses, enabling controlled interpolation of the kaon mass toward the physical regime. The second smallest kappa, $\kappa = 0.154581$, is associated with the strange quark. This value is determined by interpolating the $\rho$ meson mass data computed in Ref.~\cite{pion4p}, as illustrated in Fig.~\ref{fig:skappa}.

The interpolation is performed using an unweighted linear fit, without incorporating the statistical uncertainties of the lattice data points. This choice is justified by the smooth and nearly linear dependence of the $\rho$ meson mass on $\kappa$ over the range considered, such that the fitted result is largely insensitive to the precise weighting of the data. In particular, we have verified that performing a correlated fit, incorporating the full covariance matrix, yields a consistent interpolation with only small statistical uncertainties. Since the present analysis requires only an approximate determination of the strange quark kappa, the unweighted fit is sufficient for our purposes and provides a stable estimate of the underlying trend.

The kaon masses associated with each $\kappa$ are extracted directly from our two-point correlation functions. For every ensemble, we analyze a total of 500 statistically independent gauge configurations, ensuring adequate suppression of autocorrelations and reliable error estimation. The lattice spacing for this setup has been determined nonperturbatively in Ref.~\cite{CABASINO1991195}, yielding an inverse lattice spacing of $1/a = 2.312\ \text{GeV}$. This scale setting provides the necessary physical normalization for all dimensionful observables reported below.

We impose Dirichlet (open) boundary conditions in the temporal direction, while periodic boundary conditions are applied along each spatial axis. The kaon source is fixed at $t_0 = 7$ and the sink at $t_3 = 42$, within the full temporal extent $t \in [1,48]$. Two current insertions enter the four-point correlation function: one located at a fixed time $t_1$, and the second, $t_2$, scanned over a suitable range to extract the relevant time dependence of the Compton amplitude. Discrete lattice momenta are defined through integer triplets $\{n_x,n_y,n_z\}$,
\[
\bm q = \left( \frac{2\pi n_x}{L_x},\, \frac{2\pi n_y}{L_y},\, \frac{2\pi n_z}{L_z} \right),
\qquad n_x,n_y,n_z \in \mathbb{Z},
\]
which follow from imposing spatial periodicity. In practice, we restrict our analysis to four $\{n_x,n_y,n_z\}$ representative momentum classes,
\[
\{0,0,\pm1\},\quad \{0,\pm1,\pm1\},\quad \{0,0,\pm2\},\quad \{0,\pm2,\pm1\},
\]
where results are averaged over positive and negative momentum components. Due to the extended source in the $x$ direction, our calculation is sensitive only to momentum components in the transverse $(y,z)$ plane. The momentum classes listed above therefore correspond to the lowest nonzero momenta accessible in this framework.
\begin{figure}[h]
\centering
\includegraphics[scale=0.6]{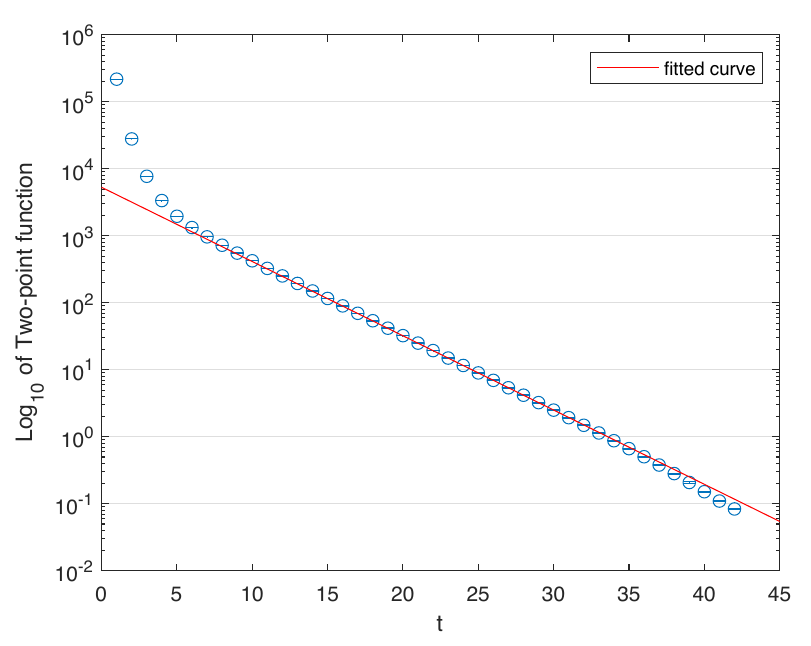} \\
\caption{Two-point correlation function for the kaon with corresponding exponential fit. 
The lattice data points are shown together with the fitted curve, from which the kaon mass is extracted.}
\label{fig:kaon_mass}
\end{figure}
%
\subsection{Raw correlation functions}

First, we need to calculate the kaon mass from the two-point function data associated with each kappa value. To obtain the kaon masses, we analyze the two-point correlation functions computed on the lattice. 
By fitting the correlator data to the expected exponential decay form, we extract the corresponding energy levels, 
with the ground-state mass identified in the large Euclidean time region. 
The fit results for the smallest mass, together with the lattice data, are displayed in Fig.~\ref{fig:kaon_mass}, where the quality of the fit 
and the consistency across time slices can be observed.
\begin{figure}[h]
\centering
\includegraphics[scale=0.8]{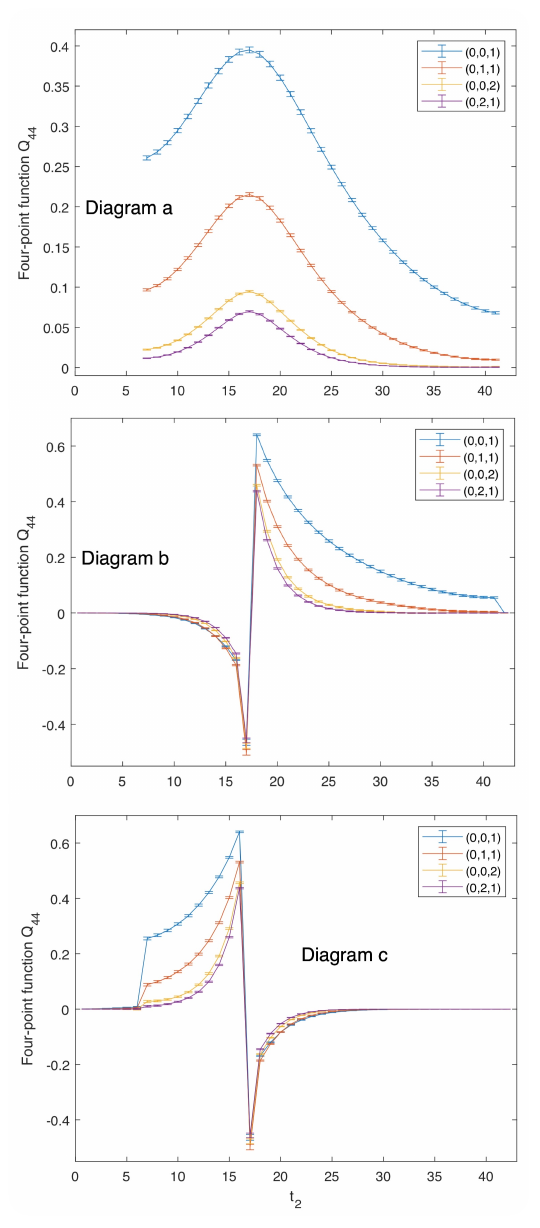}
\caption{Four-point functions from the connected diagrams, normalized to unity, as a function of current separation. The kaon walls are located at $t_0=7$ and $t_3=42$, and the fixed current insertion at $t_1=18$.}
\label{fig:Q44PS}
\end{figure}

\begin{figure}[H]
\centering
\includegraphics[scale=0.42]{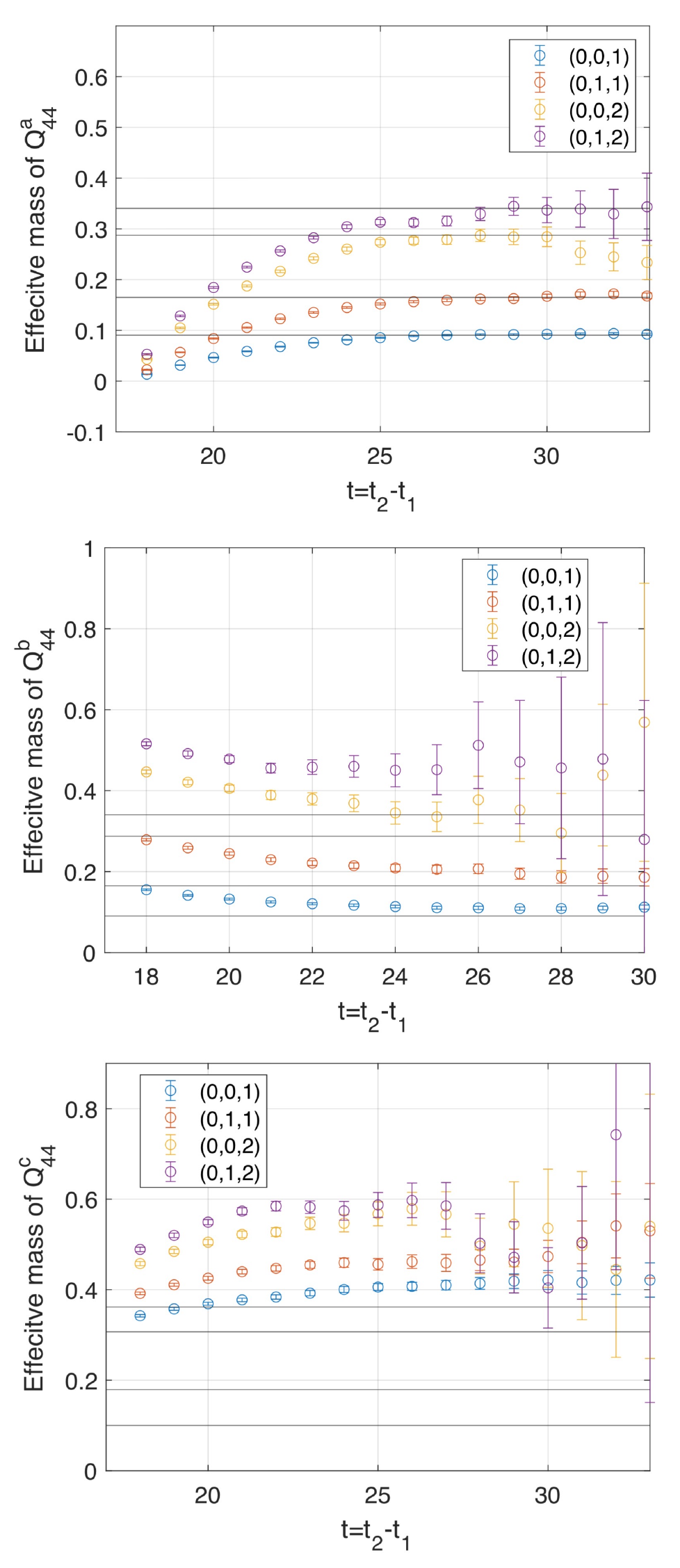}
\caption{Effective mass functions for the connected diagrams as a function of current separation. The horizontal gridlines indicate the value of $E_K-m_K$ where the continuum dispersion relation  $E_K=\sqrt{\bm q^2+m_K^2}$ is used. Diagrams (a), (b), and (c) correspond to the top, middle, and bottom plots, respectively.}
\label{fig:EMsep}
\end{figure}

Having established the kaon two-point correlation functions, we present in Fig.~\ref{fig:Q44PS} the raw, normalized four-point functions $Q_{44}$ evaluated at four distinct values of the spatial momentum $\bm q$ for a pion mass of $m_\pi = 600$~MeV. For ease of comparison, all data points for $Q_{44}$ are shown on a common linear scale. The corresponding effective mass functions, extracted from $Q_{44}$, are shown in Fig.~\ref{fig:EMsep}. The results shown include only the connected contributions, corresponding to diagrams (a), (b), and (c) in Fig.~\ref{fig:diagram-4pt1}.

A subtlety arises at the coincident insertion point $t_1 = t_2$. While this point is well behaved in diagram (a), diagrams (b) and (c) exhibit irregular behavior for all values of $\bm q$. This is associated with the presence of a contact term, as discussed in ~\cite{pion4p}. To avoid contamination from such short-distance artifacts, this point is excluded from our subsequent analysis.

Another notable feature is the symmetry observed between diagrams (b) and (c). Specifically, the data for diagram (b) at times around $t_1 = 18$ are mirror images of those for diagram (c), reflecting the fact that these diagrams correspond to the two distinct time orderings of the same underlying quark-line topology. In principle, this symmetry could be exploited to reduce the computational cost by evaluating only one of the two contributions. In the present work, however, all three diagrams are computed explicitly and combined in the interval $18 \leq t_1 \leq t_3 = 41$ to construct the final signal. All three diagrams are evaluated at the same time intervals to avoid any systematic fluctuations. We also remark that the $Q_{44}$ contribution from diagram (c) is negative definite, in contrast to the positive-definite signals observed in diagrams (a) and (b).

Finally, the effective mass extracted from $Q_{44}$ for diagram (a), as shown in Fig.~\ref{fig:EMsep}, approaches the expected asymptotic value $E_K - m_K$ at large temporal separations between $t_1$ and $t_2$. This behavior indicates that the corresponding four-point function is dominated by the elastic kaon intermediate state, with an exponential fall-off governed by $E_K - m_K$. A similar trend is observed for diagram (b), although deviations from the asymptotic behavior become more pronounced at higher momenta, likely reflecting increased excited-state contamination.

In contrast, diagram (c) exhibits markedly different behavior: its effective mass approaches values significantly larger than $E_K - m_K$, indicating dominance by inelastic contributions. This suggests that the intermediate state in this channel is not a single-kaon state but rather a higher-mass multi-quark configuration, such as a four-quark state, which contributes with a substantially larger energy gap. This qualitative difference underscores the distinct physical content encoded in the various diagram topologies and motivates their separate examination in the analysis of kaon structure observables.

\subsection{Elastic form factor}
\begin{figure}[b]
\includegraphics[scale=0.45]{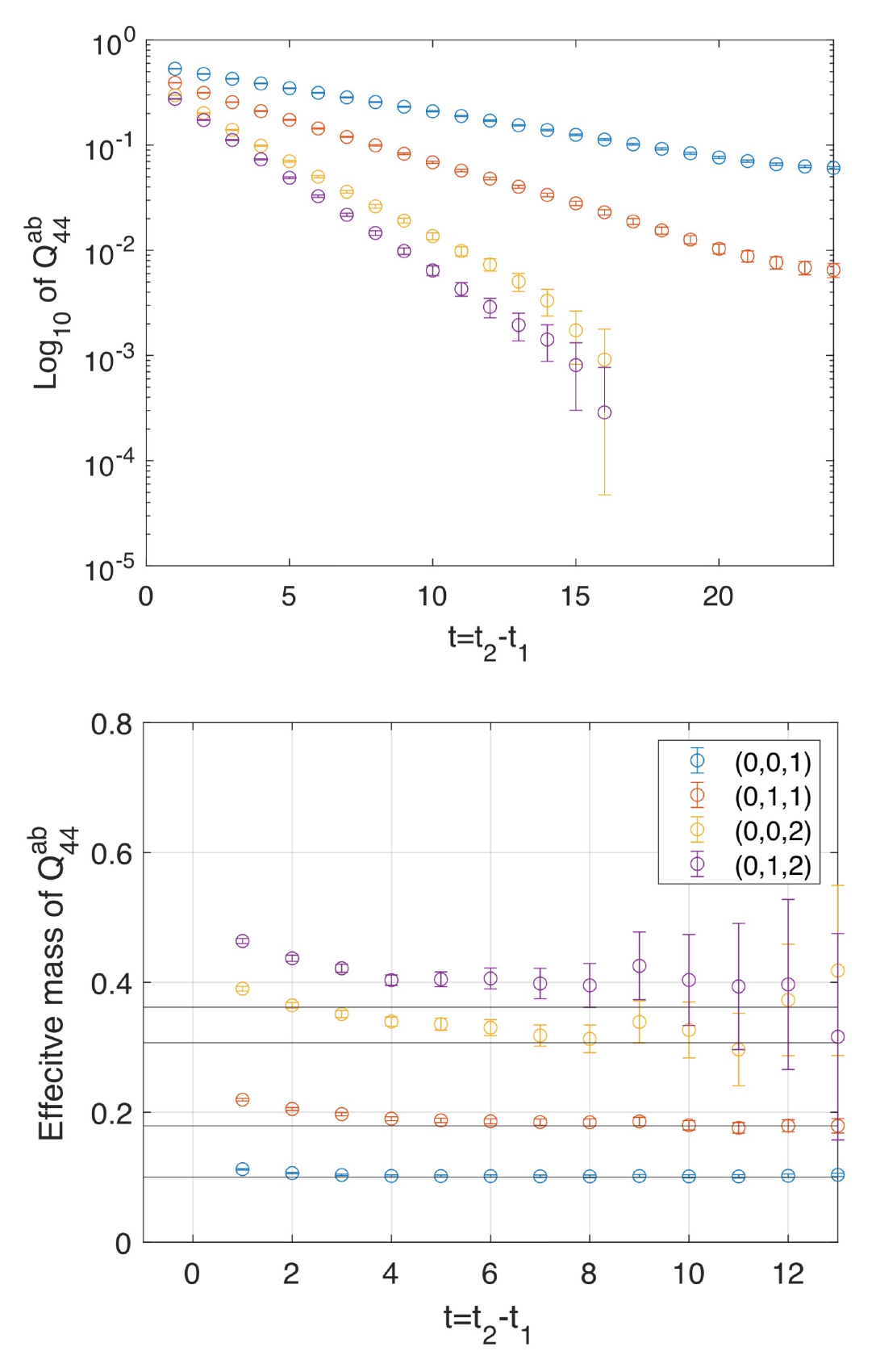}
\caption{
Normalized four-point functions from diagrams (a) and (b) in log plot and their effective mass functions at different values of $\bm q$ and $m_\pi=600$ MeV.
They are plotted as functions of  time separation $t=t_2-t_1$ between the two currents relative to fixed $t_1=18$. Negative data is omitted in the log plot.
The horizontal gridlines in the effective mass are $E_{K}-m_{K}$ using continuum dispersion relation for $E_{K}$ with measured $m_{K}$.
 These functions are used to extract the elastic contributions $Q^{elas}_{44}$.
}
\label{fig:Q44ab}
\end{figure}

The expression for the electric polarizability given in Eq.~\eqref{eq:kaon4pt} depends on two key hadronic quantities: the kaon charge radius $r_E$ and the elastic contribution to the four-point function, $Q^{\text{elas}}_{44}$. Both quantities can be extracted from the long-time behavior of the kaon four-point correlation function $Q_{44}$. As shown in Eq.~\eqref{eq:elas4pt}, the elastic contribution is expected to exhibit a single-exponential decay governed by the energy gap $E_K - m_K$, with the kaon form factor $F_K$ entering as an overall amplitude.

We observe that diagrams (a) and (b) display the expected elastic fall-off behavior, while diagram (c) is dominated by inelastic contributions and does not approach the elastic limit. Since diagram (c) contaminates the long-time behavior with higher-energy intermediate states, it can be safely omitted when isolating the elastic contribution. We therefore restrict our analysis to the combined contribution of diagrams (a) and (b), which significantly improves the stability and reliability of the form factor extraction by suppressing inelastic contamination. This restriction may be viewed as an optimization of the four-point function analysis tailored to the elastic kaon channel.

For conserved currents, the ratio of the four-point function to the two-point function at zero-momentum is expected to approach a constant charge factor that is independent of the current insertion times $t_1$ and $t_2$. For diagram (a), where the two currents couple to different valence quarks, this factor is given by $2 q_u q_{\bar{s}}$, which in the isospin-symmetric limit corresponds to $4/9$ for a charged kaon composed of a light quark and a strange antiquark. For diagrams (b) and (c), in which both currents couple to the same quark line, the expected factor is $q_u^2 + q_{\bar{s}}^2 = 5/9$. These charge factors are put in manually when the different diagrams are added together.

An illustrative example of the resulting four-point function,
$Q^{(a,b)}_{44}$, constructed from diagrams (a) and (b) only, is shown in Fig.~\ref{fig:Q44ab}, together with the corresponding effective mass. The data are plotted as a function of the current–current separation $t = t_2 - t_1$, focusing on the temporal window between the source and sink times $t_1$ and $t_3$, where the signal is well defined. As discussed earlier, the point $t=0$ is excluded from the analysis due to the presence of contact terms.

\begin{table*}[t!]
\caption{Kaon form factor $F_K(q^2)$ from four-point functions.  An example of the data to be fitted is given in Fig.~\ref{fig:Q44ab}. The fit form is in Eq.\eqref{eq:elas4pt} with $F_K$ treated as the free parameter and, $m_K$ and $E_K$ taken from the measured value. The four columns correspond to $\bm q=\{0,0,1\}, \{0,1,1\}, \{0,0,2\},  \{0,1,2\}$ from left to right.}
\label{tab:ff}
\begin{tabular}{c}
$      
\renewcommand{\arraystretch}{1.2}
\begin{array}{l|cccc}
\hline
  & \text{} & m_{\pi }\text{=800 MeV} & \text{} & \text{} \\
\hline
 F_{K} & 0.7685\pm 0.0015 & 0.6471\pm 0.0014 & 0.5433\pm 0.0021 & 0.5043\pm 0.0041 \\
 \text{Fit range} & \text{$\{$7,19$\}$} & \text{$\{$9,21$\}$} & \text{$\{$6,13$\}$} & \text{$\{$6,10$\}$} \\
 \chi ^2\text{/dof} & 0.09 & 0.10 & 0.38 & 1.22 \\
 \hline
  & \text{} & m_{\pi }\text{=750 MeV} & \text{} & \text{} \\
  \hline
 F_{K} & 0.7633\pm 0.0015 & 0.6413\pm 0.0015 & 0.5387\pm 0.0022 & 0.4998\pm 0.0044 \\
  \text{Fit range} & \text{$\{$7,19$\}$} & \text{$\{$9,21$\}$} & \text{$\{$6,13$\}$} & \text{$\{$6,10$\}$} \\
 \chi ^2\text{/dof} & 0.10 & 0.09 & 0.26 & 0.89 \\
\hline
  & \text{} & m_{\pi }\text{=600 MeV} & \text{} & \text{} \\
\hline
 F_{K} & 0.7446\pm 0.0019 & 0.6222\pm 0.0022 & 0.5225\pm 0.0027 & 0.4952\pm 0.0058 \\
 \text{Fit range} & \text{$\{$9,21$\}$} & \text{$\{$10,23$\}$} & \text{$\{$6,13$\}$} & \text{$\{$5,9$\}$} \\
 \chi ^2\text{/dof} & 0.32 & 0.25 & 0.40 & 3.20 \\
\hline
  & \text{} & m_{\pi }\text{=370 MeV} & \text{} & \text{} \\
\hline
 F_{K} & 0.7283\pm 0.0024 & 0.6124\pm 0.0024 & 0.4892\pm 0.0071 & 0.4776\pm 0.012 \\
 \text{Fit range} & \text{$\{$4,16$\}$} & \text{$\{$5,17$\}$} & \text{$\{$6,13$\}$} & \text{$\{$5,9$\}$} \\
 \chi ^2\text{/dof} & 0.31 & 0.34 & 0.71 & 2.96 \\
 \hline
\end{array}$
\end{tabular}
\end{table*}

In the intermediate-time region, the effective mass extracted from $Q^{(a,b)}_{44}$ is seen to approach the expected value $E_K - m_K$, signaling dominance by the elastic kaon intermediate state. This agreement is most pronounced at smaller values of the momentum transfer, while at larger momenta the signal deteriorates more rapidly with increasing time separation due to enhanced statistical noise. At very large times, the effects of the Dirichlet temporal boundary condition become visible, forcing the effective mass to curve slightly downward. 

In contrast to the earlier pion study~\cite{pion4p}, where both $\{F_\pi, E_\pi\}$ were treated as free parameters in the fit to $Q^{elas}_{44}$, here we fix $E_K$ to the value calculated directly 
from the measured kaon mass via the continuum dispersion relation, leaving only $F_K$ as a 
free parameter. This methodological difference means that any deviation from the continuum 
dispersion relation will manifest directly in the quality of our fits, rather than being absorbed into a free $E_K$ parameter. Details of the fits at all four pion masses are given in Table~\ref{tab:ff}.

We note that the reduced $\chi^2$ values become large for the two lowest pion masses at the highest momentum. This is not unexpected, and is likely related to the onset of failure of the 
continuum dispersion relation at higher momenta. Nevertheless, as can be seen from 
Fig.~\ref{fig:Q44ab}, the results at this momentum remain consistent with the local energy 
expectation from the two-point function, suggesting that the fits are still providing a 
reliable description of the elastic contribution despite the elevated $\chi^2$.

Once the kaon form factor data are obtained, they must be parametrized in order to extract the charge radius. A commonly used choice is the monopole form motivated by vector meson dominance. While such a parametrization can provide a qualitative description of the data, we do not pursue it further in this work. Instead, we parametrize the kaon electromagnetic form factor using the model-independent $z$-expansion~\cite{zexpansion}. This approach exploits the analytic structure of the form factor in the complex momentum-transfer plane and provides a systematically improvable expansion that is well suited to lattice data with limited momentum coverage. The $z$-expansion of the kaon form factor is written as
\beqs
& F_K(q^2) = 1 + \sum_{k=1}^{k_{\max}} a_k \, z^k , \\
& z(t;t_0) \equiv 
\frac{\sqrt{t_{\rm cut}-t} - \sqrt{t_{\rm cut}-t_0}}
     {\sqrt{t_{\rm cut}-t} + \sqrt{t_{\rm cut}-t_0}} , \\
& t = - q^2 , \qquad t_{\rm cut} = 4 m_K^2 ,
\label{eq:z}
\eeqs
where the coefficients $a_k$ are free fit parameters and $t_{\rm cut}$ corresponds to the two-kaon production threshold. We choose the expansion point $t_0 = 0$, such that the normalization condition $F_K(0)=1$ is satisfied by construction.
\begin{figure}[h]
\includegraphics[scale=0.44]{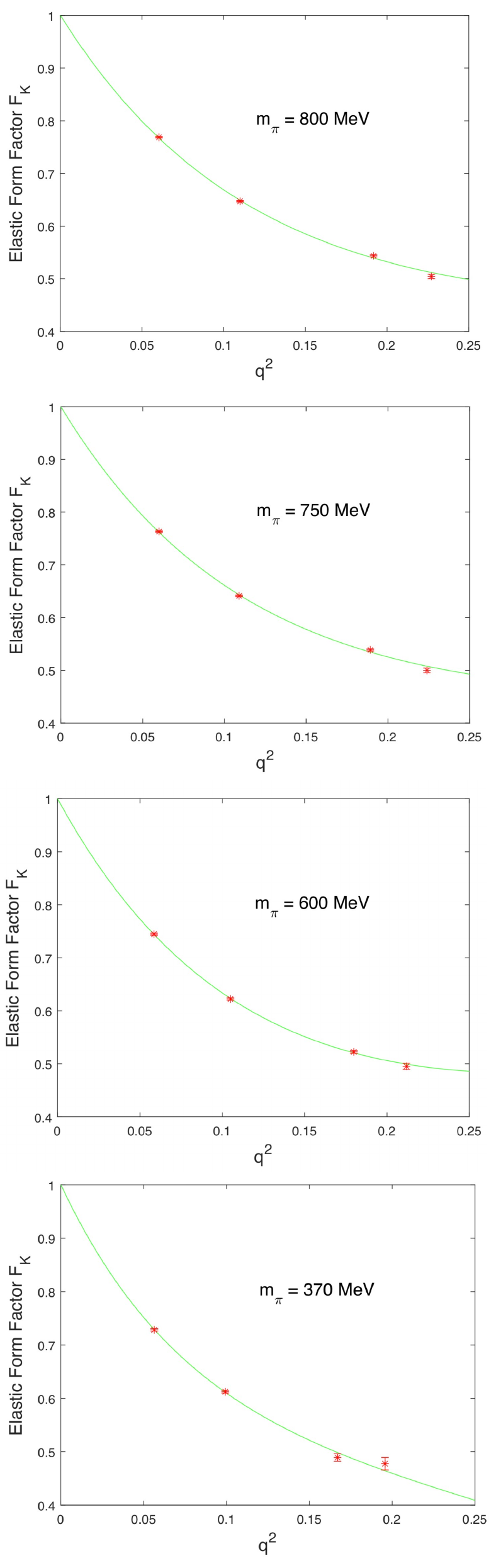}
\caption{
Kaon elastic form factors extracted from four-point functions. $q^2$ is the four-momentum squared. The red data points are the measured values in Table~\ref{tab:ff}. The green solid line is a fit to the z-expansion in Eq.~\eqref{eq:z}. 
}
\label{fig:ff}
\end{figure}

For the momentum range accessible in this study, we find that truncating the expansion at $k_{\max}=3$ is sufficient to achieve stable fits with good statistical quality. The
results are illustrated in Fig.~\ref{fig:ff}. The use of the $z$-expansion allows for a controlled determination of the low-$q^2$ behavior of the form factor and, in particular, facilitates a reliable extraction of the kaon charge radius from the slope at vanishing momentum transfer.

Using the $z$-expansion fit to the kaon form factor, the mean-square charge radius is obtained from the slope at zero-momentum transfer as defined in Eq.~(\ref{chargeradius}).
\begin{figure}[H]
\includegraphics[scale=0.6]{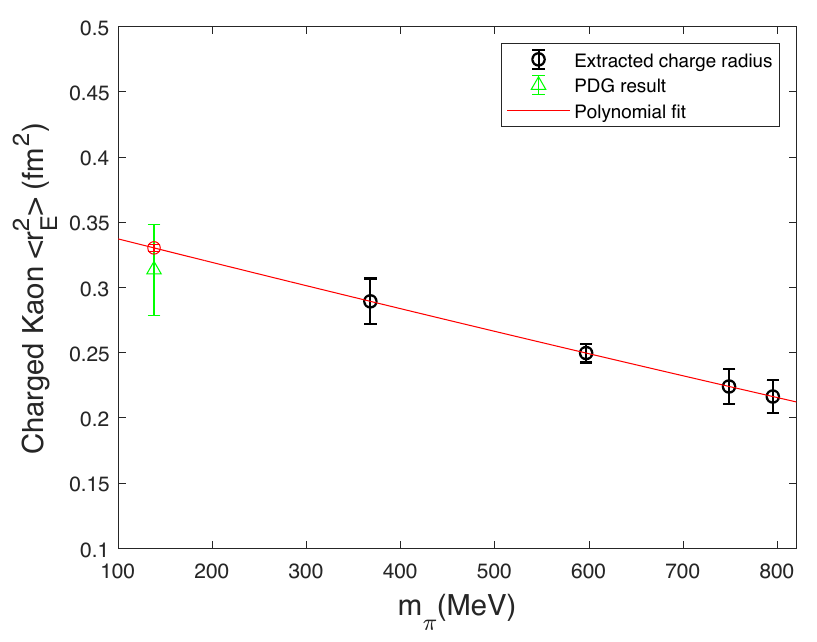}
\caption{
Chiral extrapolation of mean-square charge radius. The red point indicates the extrapolated value at the physical pion mass. Because the lightest simulated pion mass is $370~\mathrm{MeV}$, the extrapolation should be regarded as indicative rather than as a controlled physical-point determination. Green triangle is the PDG value for squared charge radius. 
}
\label{fig:chargeradius}
\end{figure}

Using the extracted charge radii, we perform a chiral extrapolation employing a quadratic ansatz of the form $a + b\, m_{\pi} + c\, m_{\pi}^2$, as shown in Fig.~\ref{fig:chargeradius}. The extrapolated charge radius at the physical point is consistent with the Particle Data Group (PDG) result. The values can be found in Table~\ref{tab:final}. We note that the error obtained from the fit is unusually small, reflecting the strong constraints imposed by the smooth behavior of the data and the use of a functional prediction interval, which quantifies the uncertainty of the fitted mean. In this sense, the quoted error should be interpreted as representative of the internal consistency of the fit rather than a complete estimate of all sources of uncertainty.
\begin{table*}[t!]
\caption{Summary of results in physical units from two-point and four-point functions. 
 $\alpha_E$ elastic and  $\alpha_E$ total are extrapolated to the physical point. The $\alpha_E$  inelastic at the physical point is taken as the difference of the two.
Known values from ChPT~\cite{Moinester:2019sew} and PDG~\cite{pdg} are listed for comparison purposes.
All polarizabilities are in units of $10^{-4}\;\text{fm}^3$.
}
\label{tab:final}
\begin{tabular}{c}
$      
\renewcommand{\arraystretch}{1.2}
\begin{array}{l|ccccc|c}
\hline
  & \text{$\kappa $=0.1543} & \text{$\kappa $=0.154581} & \text{$\kappa $=0.1555} & \text{$\kappa $=0.1565} & \text{physical point} & \text{known value}
   \\
\hline
 m_{K }\text{ (MeV)} & 772.5\pm 1.7 & 751.0\pm 1.8 & 677.1\pm 1.9 & 590.0\pm 2.4 &  & 493.6\pm0.013 \text{ (PDG)} \\
 \langl r_E^2\text{$\rangl$ (}\text{fm}^2\text{) } & 0.2164\pm 0.0126 & 0.2241\pm 0.0133 & 0.2498\pm 0.0073 & 0.2894\pm 0.0175  & 0.3303\pm 0.0028 & 0.3136\pm 0.0347 \text{ (PDG)}\\     
 \alpha _E\text{ elastic}  & 1.345\pm 0.078 & 1.433\pm 0.085 & 1.771\pm 0.052 & 2.355\pm 0.143 & 2.969\pm 0.274 & 3.0494\pm 0.0008 \\
  \alpha _E\text{ inelastic }  & -0.511\pm 0.115 & -0.533\pm 0.123 & -0.626\pm 0.124 & -0.921\pm 0.096 &  &\\
 \alpha _E\text{ total } & 0.835\pm 0.139 & 0.901\pm 0.150 & 1.146\pm 0.135 & 1.434\pm 0.172 & 1.682\pm  0.523 & 0.58 \text{ (ChPT)}\\
  \hline
  \hline
\end{array}
$   
\end{tabular}
\end{table*}
%
\subsection{Electric polarizability}

\begin{figure}[H]
\includegraphics[scale=0.5]{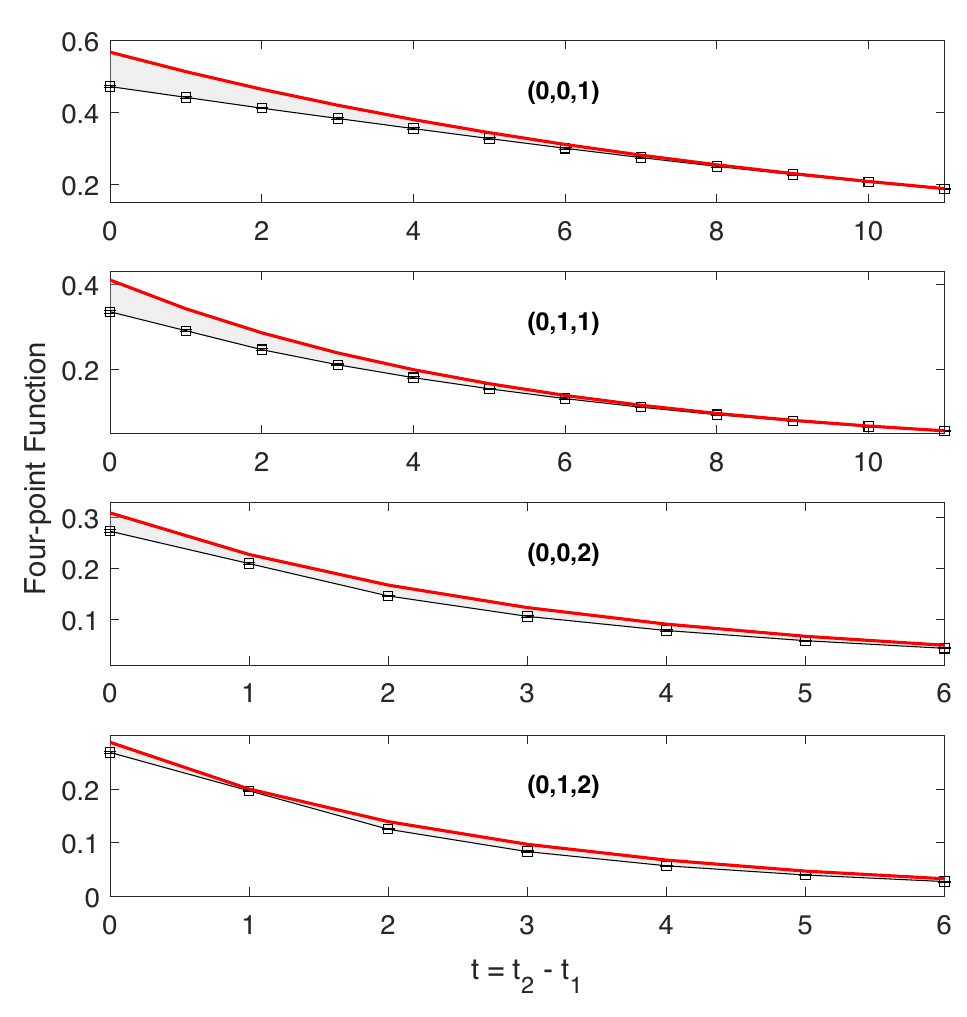}
\caption{
Total $Q_{44}$ and elastic $Q_{44}^{elas}$ at different values of $\bm q$ at $m_\pi=600$ MeV.
 The area between total $Q_{44}$ and elastic $Q_{44}^{elas}$, $(1/a)\int dt \big[ Q_{44}(\bm q,t)-Q_{44}^{elas}(\bm q,t)\big]$,  is the dimensionless signal contributing to polarizability.
}
\label{fig:QQ}
\end{figure}

Having isolated the elastic contribution $Q_{44}^{\rm elas}$, we now turn to the determination of the inelastic part of the electric polarizability $\alpha_E$ appearing in Eq.~\eqref{eq:kaon4pt}. In Fig.~\ref{fig:QQ}, we display the full four-point function $Q_{44}$, including contributions from all three connected diagrams, together with the elastic component $Q_{44}^{\rm elas}$, as functions of the current–current separation $t = t_2 - t_1$. As a representative example, we show results for a pion mass of $m_\pi = 600$~MeV; the qualitative behavior at the other pion masses studied is similar.

Although $Q_{44}^{\rm elas}$ is determined by fitting the large-time behavior of the four-point function, the subtraction $Q_{44} - Q_{44}^{\rm elas}$ is carried out over the entire temporal range using the functional form given in Eq.~\eqref{eq:elas4pt}. This procedure is essential, as the dominant contribution to the inelastic term arises from the short-time region, where inelastic intermediate states are most significant. We observe that $Q_{44}^{\rm elas}$ is consistently larger than the full correlator $Q_{44}$, indicating that the inelastic contribution to $\alpha_E$ enters with a negative sign. The time integral of the inelastic term is therefore given by the negative of the shaded area between the two curves in Fig.~\ref{fig:QQ}.

A technical subtlety arises from the inclusion of the $t=0$ point, which contains unphysical contact-term contributions in the full four-point function $Q_{44}$, as discussed earlier. In our analysis, this point is not used directly; instead, its contribution is reconstructed via a linear extrapolation of $Q_{44}$ to $t=0$ using the data at $t=1$ and $t=2$. Since the interval between $t=0$ and $t=1$ provides the largest single contribution to the integral, this procedure ensures it is retained, at the cost of a systematic uncertainty of order $O(a^2)$, with the extrapolation error itself scaling as $O(a)$. As the lattice spacing is reduced and the continuum limit is approached, this systematic effect vanishes, as the width of this interval shrinks to zero. In contrast, the elastic contribution $Q_{44}^{\rm elas}$ can be safely evaluated at $t=0$ using its analytic functional form, and therefore does not suffer from the same ambiguity.
\begin{figure}[h]
\includegraphics[scale=0.4]{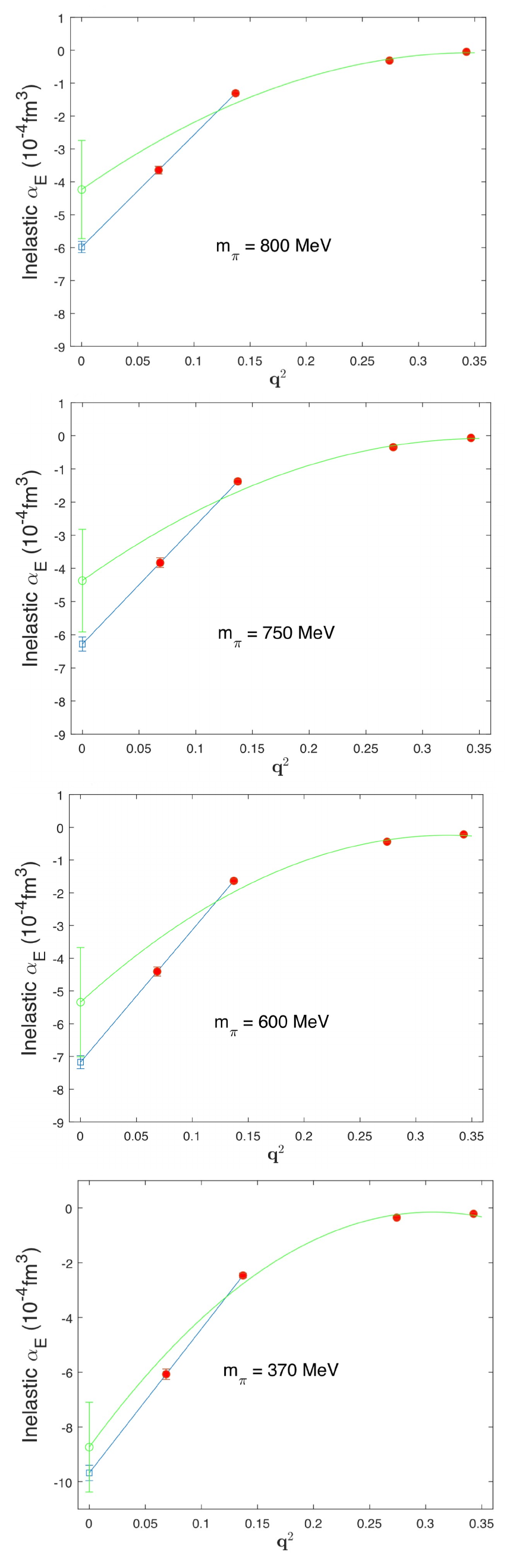}
\caption{Three-momentum dependence of the inelastic term in Eq.~\eqref{eq:kaon4pt} and its extrapolation to $\bm q^2=0$ at all pion masses.
Red points are based on the areas acquired from Fig.~\ref{fig:QQ}. Green curve is a quadratic extrapolation using all points.   Blue curve is a linear extrapolation based on the two lowest points.
}
\label{fig:Qzero}
\end{figure}
\begin{figure}[H]
\includegraphics[scale=0.6]{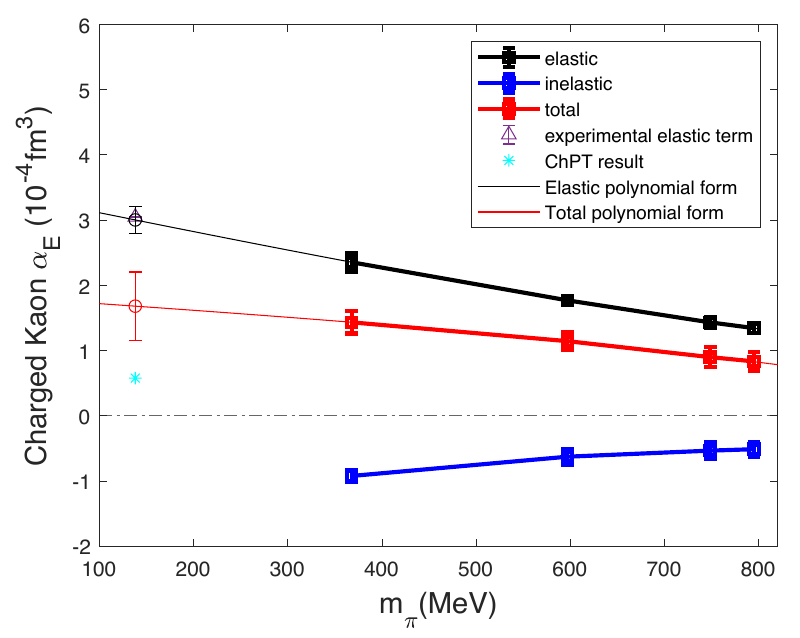}
\caption{
Pion mass dependence of electric polarizability  of a charged kaon from  four-point functions in lattice QCD.
Elastic and inelastic contributions correspond to the two terms in the formula in Eq.\eqref{eq:kaon4pt}. 
Elastic and total are extrapolated to the physical point. Inelastic is the difference of the two. Empty circles are extrapolated values at the physical point, shown as an
indicative extrapolation rather than a controlled physical-point
determination. Cyan star is known value from ChPT.
Purple triangle is the PDG value for elastic.
}
\label{fig:chiral}
\end{figure}

The inelastic contribution to the electric polarizability is then obtained by multiplying the time integral of $Q_{44} - Q_{44}^{\rm elas}$ by the kinematic factor $2\alpha / \bm q^{\,2}$, yielding a momentum-dependent quantity. Since $\alpha_E$ is a static observable, we extrapolate this result smoothly to $\bm q^2 = 0$. To assess the stability of this extrapolation, we consider two fitting strategies: a quadratic fit of the form $a + b x + c x^2$ with $x = \bm q^2$ using all four-momentum points and a linear fit using the two lowest points. The resulting extrapolations for all pion masses are shown in Fig.~\ref{fig:Qzero}. A noticeable spread among the extrapolated values is observed. We interpret this spread as a systematic uncertainty associated with the $\bm q^2 \to 0$ extrapolation. Quantitatively, we take the average of the spread among the two extrapolated values at each pion mass as the central value, with its associated statistical uncertainty. Half of the spread is then assigned as a systematic uncertainty. The statistical and systematic errors are combined in quadrature and propagated through to the final determination of $\alpha_E$. For the present data set, the statistical uncertainties are relatively small, and the extrapolation-induced systematic uncertainties dominate the error budget of the inelastic contribution.

Finally, the elastic and inelastic components are combined according to Eq.~\eqref{eq:kaon4pt} to obtain the total electric polarizability $\alpha_E$ in physical units. To explore the behavior of $\alpha_E$ toward smaller pion masses and ultimately toward the physical point, we perform a smooth mass extrapolation using the results obtained at the four simulated pion masses. Given that the simulated masses remain relatively heavy, we employ a simple polynomial ansatz $a + b\, m_K + c\, m_K^3$ to describe the mass dependence. Chiral expansions based on SU(3) chiral perturbation theory are formally valid only in the low-energy regime, where the typical scales are well below the mass of the first resonances~\cite{Gonzalez-Solis2018-yo}. At higher energies or masses, the convergence of the chiral expansion deteriorates and additional resonance contributions must be included explicitly. In the present case, the simulated pion masses lie outside the regime where a controlled chiral expansion is expected to hold, and we therefore adopt a polynomial parametrization, which provides a stable and model-independent description suitable for this proof-of-concept study. Accordingly, we regard the results at the simulated masses, summarized in
Table~\ref{tab:final}, as the primary numerical output of this study; the
physical-point values are provided for qualitative comparison and should be understood
as an extrapolation rather than a controlled determination, since the
lightest available pion mass ($370~\mathrm{MeV}$) lies well outside the
regime of a controlled chiral expansion.

The final results are summarized in Fig.~\ref{fig:chiral} and Table~\ref{tab:final}. Across the range of kaon masses studied, we observe a clear and consistent pattern: the elastic contribution to the electric polarizability is positive, while the inelastic contribution is negative and smaller in magnitude. Their partial cancellation yields a positive total value of $\alpha_E$, and this behavior persists through the physical-point extrapolation. Quantitatively, we find the elastic contribution to be $\alpha_E^\text{elastic} = 2.969 \pm 0.274$, in good agreement with the experimental value $3.0494 \pm 0.0008$ obtained from the PDG values for the charged kaon charge radius and mass. The total electric polarizability, including both elastic and inelastic
contributions, is $\alpha_E = (1.682\pm  0.523)$ at the extrapolated physical
point. This is of the same sign and order of magnitude as the SU(3) chiral
perturbation theory value of $0.58$\cite{Moinester:2019sew}. We emphasize, however, that the quoted uncertainty accounts only for statistical and extrapolation errors. Systematic effects arising from quenching, disconnected diagrams, finite-volume artifacts, and the use of a single lattice spacing have not been quantified. Therefore, the observed agreement should be viewed as encouraging rather than as a definitive quantitative test of ChPT.

\section{Summary and outlook}
\label{sec:con}
In this work we have presented a lattice QCD study of the charged kaon electric polarizability using a four-point function formulation of Euclidean Compton scattering~\cite{4point}. This calculation extends the four-point function methodology previously developed for the pion to a meson containing a strange quark and provides a nontrivial test of the approach in a system with a different internal structure and mass scale.

A key methodological distinction from the charged pion study is the use of kinematics that allow the four-point function to be interpreted in terms of a spatial charge-density distribution in the transverse $(y,z)$ plane. By injecting momentum only in the transverse directions, the resulting Compton tensor admits a representation that isolates the response of the hadron to an external electric field in transverse coordinate space. This formulation offers a complementary perspective to the time-integrated approach employed in the pion calculation. In addition, momentum reconstruction is deferred entirely to the analysis stage, with no explicit momentum projection imposed during propagator construction. This contrasts with approaches in which momenta are fixed a priori, and has the advantage of allowing all injected momenta to be reconstructed within a single dataset.

The charged kaon electric polarizability was decomposed into elastic and inelastic contributions. The elastic term was determined from the kaon electromagnetic form factor, which was extracted from the long-time behavior of the four-point function. A model-independent $z$-expansion was used to parametrize the form factor, yielding the kaon charge radius and thereby fixing the elastic contribution. The inelastic contribution was obtained by subtracting the elastic component from the full four-point function and integrating over Euclidean time, followed by an extrapolation to the static $\bm q^2 \to 0$ limit.

An important feature observed in the kaon four-point functions is the distinct behavior of the different quark-line topologies. Diagrams (a) and (b) are dominated at large Euclidean times by the elastic kaon intermediate state and therefore provide clean access to the electromagnetic form factor. In contrast, diagram (c) exhibits a significantly larger effective mass, indicating dominance by higher-energy states. This separation enables a controlled isolation of elastic and inelastic dynamics and illustrates how different physical contributions are encoded in the four-point correlator.

The present calculation was carried out in the quenched approximation, at unphysically heavy pion masses, and includes only connected contributions. As such, the results should be regarded as a proof-of-principle demonstration rather than a precision determination of the kaon electric polarizability. Nevertheless, the consistency of the form factor extraction, the stability of the elastic subtraction, and the clear signal in the inelastic component indicate that the four-point function framework remains robust when applied to strange mesons.

Several extensions of this work suggest themselves. The inclusion of disconnected diagrams is a necessary step toward a complete determination of polarizabilities. It is expected to be particularly important for neutral kaons, as is seen in the neutral pion study~\cite{neutralpion}, but will contribute also in the charged case. Diagram (d) from Fig.~\ref{fig:diagram-4pt1} would be the first place to look for additional significant contributions. This diagram simply represents the numerical configuration correlation of separate (zero-momentum) pseudoscalar and (nonzero-momentum) scalar (electric case) or vector (magnetic case) two-point functions. Our initial investigation on the same lattices used here has revealed that obtaining a good signal for this contribution is possible, but challenging. In addition, an application of the present methodology to the neutral kaon, for which the elastic contribution vanishes, would provide a clean probe of purely inelastic dynamics. Finally, simulations with dynamical fermions, lighter quark masses, and larger volumes will be required to control systematic effects such as quenching, finite-volume corrections, and discretization errors. A preliminary study using dynamic fermions with normalized hypercubic (nHYP) smearing for the charged pion is currently underway and was presented at the Lattice 2025 conference~\cite{Luke:2026wkn}.

More broadly, the FR transverse-momentum formulation employed here can be extended to other hadronic observables including structure functions using the hadronic tensor approach, as well as to generalized polarizabilities accessed through nonforward kinematics. Together with the earlier pion study, this work demonstrates that four-point function methods provide a flexible and physically transparent framework for computing hadron properties directly from lattice QCD.

\vspace*{5mm}
\begin{acknowledgments}
This work was supported in part by U.S. Department of Energy under  Grant~No.~DE-FG02-95ER40907 (F.~X.~L.). W.~W. would like to acknowledge support from the Baylor College of Arts and Sciences SRA program. We are grateful to Randy Lewis for his lattice QCD matrix inverter software. The authors acknowledge the Texas Advanced Computing Center (TACC) at The University of Texas at Austin for providing computational resources that have contributed to the research results reported within this paper.
\end{acknowledgments}


\bibliography{x4ptfun}



\end{document}